\documentclass[journal]{IEEEtran}
\usepackage{cite}
\usepackage{amsmath}
\usepackage{amssymb}
\usepackage{array}
\usepackage{booktabs}
\usepackage{graphicx}
\usepackage{float}
\usepackage{subfig}
\usepackage{color}
\usepackage{amsfonts}
\usepackage{microtype}
\usepackage{algorithm}
\usepackage{algpseudocode}
\makeatletter
\renewcommand\fs@ruled{\def\@fs@cfont{\bfseries}\let\@fs@capt\floatc@ruled
	\def\@fs@pre{{\color{black}\hrule height.8pt depth0pt}\kern2pt}%
	\def\@fs@post{\kern2pt{\color{black}\hrule}\relax}%
	\def\@fs@mid{\kern2pt{\color{black}\hrule}\kern2pt}%
	\let\@fs@iftopcapt\iftrue}
\floatstyle{ruled}
\restylefloat{algorithm}
\makeatother
\begin{document}

	\title{UW-OCDM for Low-Altitude UAV Communication and Cooperative Sensing}
	\author{
		Yi~Tao,
		Zhen~Gao,~\IEEEmembership{Senior Member,~IEEE},
		Ziwei~Wan, 
		Yuezu~Lv,~\IEEEmembership{Senior Member,~IEEE},
		Hua~Wang,~\IEEEmembership{Member,~IEEE},
		Kaibin~Huang,~\IEEEmembership{Fellow,~IEEE},
		and Sheng~Chen,~\IEEEmembership{Life Fellow,~IEEE}
		\thanks{}
	}
	
	\maketitle
	
	\begin{abstract}
		Integrated sensing and communications (ISAC) is a key enabler for uncrewed aerial vehicles (UAVs) in the low-altitude economy.
		This paper proposes an ISAC waveform that embeds a unique word (UW) into orthogonal chirp division multiplexing (OCDM), termed UW-OCDM, together with corresponding communication reception and cooperative sensing schemes for high-mobility UAV scenarios.
		For communication, the embedded UW enables timing synchronization and Doppler estimation and compensation without requiring a separate synchronization sequence.
		A sparse spatio-temporal channel estimation method exploits the common channel support across multiple receive antennas and consecutive UW observations to support reliable data demodulation.
		For sensing, the deterministic UW serves as a shared prior that allows distributed base stations to construct sensing dictionaries locally without exchanging random payload symbols in real time.
		A hierarchical multi-target detection and tracking algorithm integrates direct-path interference suppression, kinematic prediction, multi-candidate screening, off-grid refinement, residual verification, and successive interference cancellation for robust localization with reduced search complexity.
		Simulation results demonstrate reliable communication and localization in highly dynamic UAV scenarios, while the proposed framework retains low-complexity frequency-domain equalization and reduces transmit-reference sharing overhead and multi-static localization complexity.
	\end{abstract}
	\vspace{-5mm}
	\section{Introduction}
	With advances in fifth-generation (5G) connectivity and low-altitude networking, the low-altitude economy has developed rapidly, with uncrewed aerial vehicles (UAVs) widely deployed in logistics, rescue operations, environmental monitoring, and as aerial base stations (BSs) \cite{tao2025multimodal}. 
	As nodes for the upcoming sixth-generation (6G) space-air-ground integrated network \cite{liu2024near,liu2025near}, UAVs leverage deployment flexibility and line-of-sight (LoS) advantages \cite{UAVLOS} to extend three-dimensional (3D) coverage and mitigate ground-network coverage blind spots.
	However, their high-speed mobility introduces severe Doppler shifts and fast time-varying channels, which not only challenge data transmission but also complicate accurate sensing and localization in low-altitude networks \cite{fastvarying}.

	This dual demand on communication and sensing makes integrated sensing and communications (ISAC) a natural framework \cite{isac2,meng2022uav,li2025iotjuav} for highly dynamic UAV scenarios. Recently, ISAC has emerged as a major focus in both academia and industry \cite{isac1,li2026cje}, aiming to achieve the software and hardware integration of communication and sensing while significantly enhancing spectral efficiency and enabling novel services. In such ground-to-air (GA) scenarios, UAVs are required to support reliable communication and high-precision sensing in an integrated manner.
	However, severe doubly-selective fading shortens channel coherence time, destroys waveform orthogonality, and triggers inter-carrier interference, thereby degrading communication and sensing performance \cite{doublefading,OCDMDOUBLE}. In the face of rapidly varying doubly-selective channels, independent signal designs for synchronization, channel estimation (CE), demodulation, and localization would not only incur high reference signal overhead but also make the performance of each module difficult to guarantee. Consequently, UAV networks urgently require a waveform architecture or signal prior that can be shared across multiple tasks.
	
	Within such a framework, waveform design is fundamental, since the performance of synchronization, CE, and subsequent sensing depends on the underlying signal structure. Several advanced waveforms have therefore been proposed \cite{li2025chirpdd}.
	Specifically, orthogonal time frequency space (OTFS) \cite{OTFS3,otfs4,li2021cross} achieves full diversity in the delay-Doppler domain, and its extension, orthogonal delay-Doppler multiplexing (ODDM) \cite{oddm}, further introduces strictly orthogonal basis functions. 
	As a parallel alternative, affine frequency division multiplexing (AFDM) \cite{Bemani2021AFDM} employs tunable chirp parameters via the discrete affine Fourier transform (DAFT). 
	OTFS and ODDM require higher-complexity two-dimensional delay-Doppler detection or equalization, whereas AFDM requires appropriate affine parameter configuration \cite{Bemani2021AFDM}.
	As another candidate, orthogonal chirp division multiplexing (OCDM) \cite{ocdm} utilizes the discrete Fresnel transform (DFnT) to generate a chirp-based eigenbasis while retaining orthogonal frequency division multiplexing (OFDM)-like low equalization complexity. 
	More importantly, the quadratic-phase chirp structure of OCDM spreads information symbols over orthogonal chirp waveforms with time-varying instantaneous frequencies. This chirp-domain spreading enhances robustness against Doppler-induced interference and helps maintain reliable transmission under the severe time variation encountered in high-mobility UAV channels \cite{OCDMDOUBLE}.
	Although existing studies demonstrate the feasibility of OCDM-based ISAC \cite{ocdmisacwcl}, its systematic integration with synchronization, CE, and cooperative sensing under severe doubly selective UAV channels remains insufficiently studied.
	
	Beyond the core waveform, the guard interval (GI) design also plays a pivotal role in enabling joint communication and sensing. 
	Early zero-padding \cite{zeropadding} avoids inter-symbol interference (ISI), but its guard interval provides no known reference for receiver processing. 
	The widely adopted cyclic prefix (CP) simplifies frequency-domain equalization (FDE) via data redundancy \cite{Muquet2002CPZP}, yet provides limited direct support for synchronization, CE, or sensing. 
	Time-domain synchronous techniques use a known pseudo-random GI for synchronization and CE, but disrupt the cyclic structure and introduce data/sequence interference requiring iterative cancellation \cite{ying2023jsac,zhou2023twc}. 
	While the chirp periodic prefix aligns naturally with ISAC due to its interference resilience \cite{CPP}, it requires appropriate configuration of prefix length and chirp parameters. 
	In contrast, the unique word (UW) design \cite{UW-OFDM} preserves the circular convolution structure required for low-complexity FDE, while its deterministic nature and favorable correlation properties make it attractive for synchronization, Doppler tracking, and CE. 
	This also makes the UW a promising shared prior for cooperative sensing in highly dynamic environments.
		
	Building on this shared prior, the communication waveform itself can be further exploited for high-precision localization in ISAC systems \cite{li2025FTN}.
	While early monostatic architectures suffer from limited accuracy and weak anti-interference capability \cite{singleBS}, multi-static cooperative schemes significantly improve localization reliability and coverage \cite{multiBS}.
	Among algorithms, conventional time difference of arrival (TDOA) localization \cite{9748989} relies on accurate inter-station synchronization \cite{wang2020tdoa}, whereas angle of arrival (AOA) localization requires reliable angle estimation and array-orientation knowledge \cite{9146923}.
	To avoid explicit TDOA/AOA extraction and fusion, compressive sensing (CS) exploits target sparsity to achieve high-resolution and noise-resilient localization \cite{cslocalization}.
	However, payload symbols are random and not naturally available as a common sensing reference at spatially separated BSs, so data-bearing multi-static sensing may require additional signaling for consistent reference reconstruction \cite{zhang2021}. 
	In contrast, the deterministic UW is known a priori and locally reconstructible at all BSs, enabling lightweight cooperative localization without exchanging instantaneous payload symbols \cite{uwlocal}.
	
	\vspace{-2mm}
	\subsection{Contributions}
	To provide reliable communication and high-precision sensing for low-altitude networks, we propose a UW-OCDM ISAC framework.
	The UW supports synchronization, Doppler compensation, and CE for communication reception, while serving as a deterministic shared reference for multi-static cooperative sensing with reduced transmit-reference sharing overhead.
	The main contributions are summarized as follows.
	
	\begin{itemize}
		\item{
			\textbf{UW-OCDM ISAC Waveform Design:}
		For UAV networks, we propose a novel ISAC waveform by embedding a deterministic UW sequence into OCDM symbols. 
		Owing to its quadratic-phase chirp structure, OCDM provides improved Doppler robustness over conventional OFDM in high-mobility channels.
		Meanwhile, the deterministic UW serves as a shared signal reference that can be reused across multiple communication and sensing operations within the proposed framework.
		}
		
		\item{
			\textbf{UW-Assisted Time Synchronization, Doppler Compensation, CE, and FDE:}
			Building on the proposed waveform, we develop a UAV receiver framework to overcome severe doubly-selective fading.
			The UW first enables timing synchronization via sliding cross-correlation. 
			The phase differences between separated ISI-free UW segments are then jointly exploited to estimate and compensate for the dominant Doppler shift, completing the time-frequency alignment. 
			Furthermore, by exploiting the spatio-temporal (ST) joint sparsity of multiple-input multiple-output (MIMO) GA channels, we formulate a multiple measurement vector (MMV) CS model. 
			A joint CE method employing MMV-based orthogonal matching pursuit (MMV-OMP) estimates the channel impulse response (CIR) for each transmit-receive link, while the UW preserves the circular convolution structure that enables low-complexity FDE and data recovery.
		}
		
		\item{
			\textbf{Multi-Static Cooperative Localization for UAV Sensing:}
			We further propose a low-overhead cooperative localization architecture across multiple BSs. 
			Distributed BSs locally construct sensing dictionaries using the known UW sequence and system parameters without real-time sharing of payload-dependent transmit references, thereby reducing real-time transmit-reference sharing overhead.
			Moreover, after direct-path interference (DPI) suppression, we develop a robust off-grid hierarchical UAV detection and tracking algorithm.
			By integrating kinematic prediction, non-maximum suppression (NMS)-based multi-candidate coarse screening, adaptive region-of-interest (ROI) expansion, off-grid refinement, residual verification, and successive interference cancellation (SIC), the proposed localization method improves robustness while reducing search complexity.
		}
	\end{itemize}

	\emph{Notations:} Boldface lowercase and uppercase letters denote column vectors and matrices, respectively, while calligraphic letters denote sets. $\mathbb{R}$ and $\mathbb{C}$ are the real and complex domains, and $\jmath=\sqrt{-1}$. $(\cdot)^T$, $(\cdot)^*$, $(\cdot)^H$, and $(\cdot)^{-1}$ denote the transpose, conjugate, conjugate transpose, and inverse, respectively. A hat denotes an estimate, while $(\cdot)^\star$ denotes an optimizer. $\langle a \rangle_2$ denotes the remainder of $a$ modulo $2$. $\operatorname{rect}(x)$ equals $1$ for $x\in[0,1)$ and $0$ otherwise. $|\mathcal{S}|$ denotes the cardinality of set $\mathcal{S}$, and $|x|$ denotes the magnitude of scalar $x$. For a vector, $|\cdot|$ and $\angle(\cdot)$ operate elementwise. For indexing, $\mathbf{a}[n]$ and $\mathbf{A}[m,n]$ are the $n$-th and $(m,n)$-th entries of vector $\mathbf{a}$ and matrix $\mathbf{A}$, respectively. $\mathbf{A}[\mathcal{I},:]$ and $\mathbf{A}[:,\mathcal{I}]$ denote the submatrices with rows and columns indexed by $\mathcal{I}$, respectively, and $\mathbf{a}[\mathcal{I}]$ is the corresponding subvector. $\|\cdot\|_2$ is the $\ell_2$ norm. $\mathbf{I}_N$ is the $N\times N$ identity matrix. $\mathbf{F}_N$ is the normalized unitary $N\times N$ discrete Fourier transform (DFT) matrix, so $\mathbf{F}_N^{-1}=\mathbf{F}_N^H$. $\delta[\cdot]$ is the Kronecker delta. $\mathcal{N}(\mu,\sigma^2)$ and $\mathcal{CN}(\mu,\sigma^2)$ denote the real Gaussian and circularly symmetric complex Gaussian distributions, respectively. $\operatorname{Re}\{\cdot\}$ extracts the real part, and $\operatorname{diag}\{\cdot\}$ constructs a diagonal matrix. {$c$ and $\lambda$ denote the speed of light and carrier wavelength, respectively.} $\odot$ and $\otimes$ denote the Hadamard and Kronecker products, respectively. $\mathcal{O}(\cdot)$ is the big-O notation.
	
	\vspace{-3mm}
	\section{System Model}
	\begin{figure}[!t]
		\captionsetup{font={footnotesize}, name = {Fig.}, labelsep = period}
		\centering
		\includegraphics[width=7.8cm]{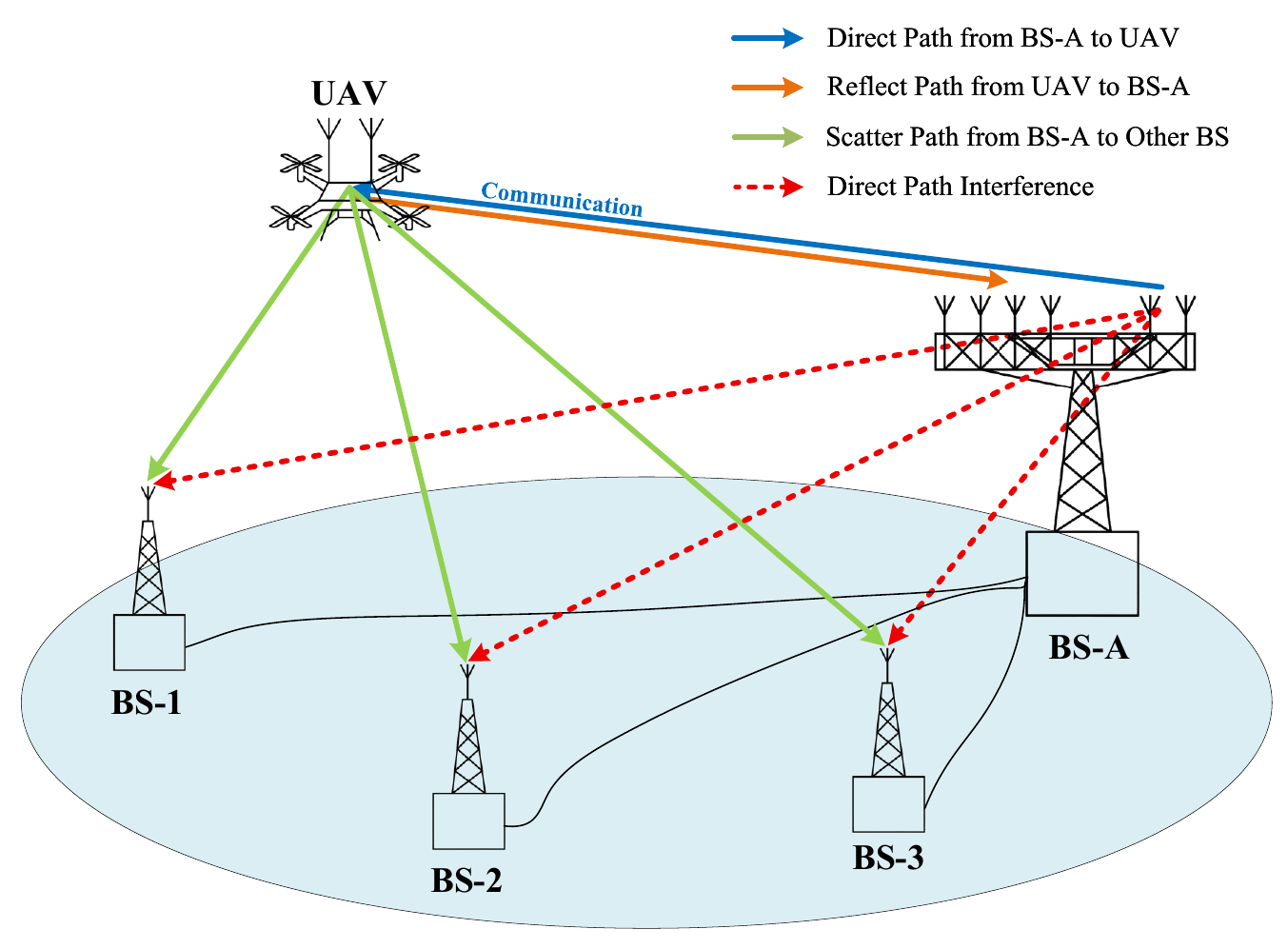}
		\caption{Illustration of the proposed joint communication and multi-static cooperative sensing for the low-altitude UAV.}
		\label{Fig_sensing}
		\vspace*{-5mm}
	\end{figure}
	
As shown in Fig.~\ref{Fig_sensing}, the primary BS (BS-A) transmits to a communication UAV, which may also act as a sensing target, while BS-A and distributed BSs capture the UAV echoes for cooperative sensing.
In this section, we first describe the doubly-selective MIMO GA downlink channel, then construct the corresponding multi-static sensing architecture.
	
	\vspace{-3mm}
	\subsection{Doubly Selective MIMO Ground-to-Air Channel Modeling}
	To fully capture the Doppler-induced fast time-variation and multipath-induced frequency selectivity in the highly dynamic UAV downlink \cite{fastvarying} \cite{wang2026jstsp} (i.e., the blue BS-A-to-UAV link in Fig.~\ref{Fig_sensing}), this subsection establishes a doubly-selective MIMO GA channel model. 
	Consider a system where the BS-A is equipped with $N_{\rm{tx}}^{(\rm{A})}$ transmit antennas and the UAV is equipped with $N_{\rm{rx}}^{(\rm{U})}$ receive antennas. 
	Let $x_t[n_{\rm{abs}}]$ denote the sample at absolute index $n_{\rm{abs}}$ of the continuous discrete time-domain ISAC transmit stream from the $t$-th antenna of BS-A, where $t \in \{1,\dots,N_{\rm{tx}}^{(\rm{A})}\}$.
	Following digital-to-analog conversion and wireless propagation, the signal captured by the $r$-th UAV antenna for $r \in \{1,\dots,N_{\rm{rx}}^{(\rm{U})}\}$ is formulated as
	\begin{equation}
		\label{yrn}
		y_r[n_{\rm{abs}}] = \sum_{t=1}^{N_{\rm{tx}}^{(\rm{A})}} \sum_{l=0}^{L_{\max}-1} h_{r,t}[n_{\rm{abs}}, l] x_{t}[n_{\rm{abs}}-l] + w_r[n_{\rm{abs}}],
	\end{equation}
	where $h_{r,t}[n_{\rm{abs}}, l]$ denotes the time-varying discrete CIR from the $t$-th transmit antenna to the $r$-th receive antenna at $n_{\rm{abs}}$ and delay tap $l \in \{0,\dots,L_{\max}-1\}$. 
	Here, $L_{\max}$ represents the maximum number of delay taps, and $w_r[n_{\rm{abs}}]\sim\mathcal{CN}(0,\sigma_w^2)$ is the additive white Gaussian noise.

	Since typical GA scenarios feature a dominant LoS path and weaker non-line-of-sight (NLoS) components \cite{10224317,liao2021jsac_aero,10195164}, we model the MIMO GA channel as a wideband Rician fading process. 
	Specifically, $\mathbf{H}[n_{\rm{abs}}, l] \in \mathbb{C}^{N_{\rm{rx}}^{(\rm{U})} \times N_{\rm{tx}}^{(\rm{A})}}$ is the complete time-varying discrete baseband CIR matrix, where its $(r, t)$-th element is $h_{r,t}[n_{\rm{abs}}, l]$. 
	With both the transmitter and receiver equipped with uniform linear arrays (ULAs), the LoS-dominated propagation characteristics \cite{fadingchannel,ke2020tsp} allow $\mathbf{H}[n_{\rm{abs}}, l]$ to be modeled as \begin{equation}
		\begin{aligned}
		\mathbf{H}[n_{\rm{abs}}, l]
		&= \sqrt{\frac{K_{\rm{Ri}}}{K_{\rm{Ri}}+1}} \mathbf{H}_{\rm{LoS}}[n_{\rm{abs}}, l] \\
		&\quad + \sqrt{\frac{1}{K_{\rm{Ri}}+1}} \mathbf{H}_{\rm{NLoS}}[n_{\rm{abs}}, l],
		\end{aligned}
	\end{equation}
	where $K_{\rm{Ri}}$ denotes the Rician K-factor.
	The deterministic LoS matrix component is expressed as
	\begin{equation}
		\begin{aligned}
		\mathbf{H}_{\rm{LoS}}[n_{\rm{abs}}, l]
		&= \mathbf{a}_{\rm{rx}}^{(\rm{U})}(\theta_{\rm{rx}}^{\rm{LoS}}) \left(\mathbf{a}_{\rm{tx}}^{(\rm{A})}(\theta_{\rm{tx}}^{\rm{LoS}})\right)^H \\
		&\quad \times e^{\jmath 2\pi f_{\rm{D,LoS}} n_{\rm{abs}} T_{\rm{s}}} \delta[l],
		\end{aligned}
	\end{equation}
	where $\mathbf{a}_{\rm{rx}}^{(\rm{U})}(\cdot) \in \mathbb{C}^{N_{\rm{rx}}^{(\rm{U})} \times 1}$ and $\mathbf{a}_{\rm{tx}}^{(\rm{A})}(\cdot) \in \mathbb{C}^{N_{\rm{tx}}^{(\rm{A})} \times 1}$ denote the steering vectors at the UAV receiver and BS-A transmitter, respectively.
	Here, $\theta_{\rm{rx}}^{\rm{LoS}}$ and $\theta_{\rm{tx}}^{\rm{LoS}}$ denote the LoS angles of arrival and departure, respectively, while $\delta[l]$ sets its delay to zero. $f_{\rm{D,LoS}}$ represents the Doppler shift associated with the LoS path, and $T_{\rm{s}}$ is the discrete sampling interval.
	Thus, the LoS term captures the spatial response and Doppler evolution.
	The NLoS component, constituted by $N_{\rm{path}}$ sparse scattering paths, can be modeled as
	\begin{equation}
		\begin{aligned}
		\mathbf{H}_{\rm{NLoS}}[n_{\rm{abs}}, l]
		&= \sum_{\iota=1}^{N_{\rm{path}}} \beta_{\iota} \mathbf{a}_{\rm{rx}}^{(\rm{U})}(\theta^{\rm{rx}}_{\iota})
		\left(\mathbf{a}_{\rm{tx}}^{(\rm{A})}(\theta^{\rm{tx}}_{\iota})\right)^H \\
		&\quad \times e^{\jmath 2\pi f^{\rm{D}}_{\iota} n_{\rm{abs}} T_{\rm{s}}} \delta[l-l_{\iota}],
		\end{aligned}
	\end{equation}
	where $\theta_{\iota}^{\rm{rx}}$ and $\theta_{\iota}^{\rm{tx}}$ are the arrival and departure angles of path $\iota$, respectively, and $\beta_{\iota} \sim \mathcal{CN}(0, \sigma_{\iota}^2)$ denotes the complex Gaussian random fading gain with $\sigma_{\iota}^2$ representing its average power.
	Furthermore, $f^{\rm{D}}_{\iota}$ represents the Doppler shift induced by the $\iota$-th NLoS path, and $l_{\iota} \in \{0,\dots, L_{\max}-1\}$ is the discrete delay index associated with the $\iota$-th NLoS path.
	
	\vspace{-3mm}
	\subsection{Multi-Static Cooperative Sensing Model}
	As illustrated in Fig.~\ref{Fig_sensing}, BS-A transmits a downlink signal that also illuminates the UAV targets \cite{DOPPLER,tao2026iotj}. The communication UAV may be included among these targets, while the remaining sensed UAVs need not be communication terminals. Their scattered echoes are received by BS-A, equipped with an $N_{\rm{rx}}^{(\rm{A})}$-element ULA, and $I$ distributed receiving BSs, each using a single omnidirectional antenna.
	Let $i \in \{1,\dots,I\}$ index the distributed receiving BSs and define $\mathcal I=\{\mathrm{A},1,\dots,I\}$, where $i'\in\mathcal I$ indexes any sensing BS. For the $q$-th UAV target with $q\in\{1,\dots,Q\}$, let $\mathbf p_q=[x_q,y_q,z_q]^T$ denote its three-dimensional Cartesian position, where $x_q$, $y_q$, and $z_q$ are its coordinates along the three spatial axes. $\mathbf p_{\mathrm{A}}$ and $\mathbf p_i$ denote the coordinates of BS-A and distributed BS $i$, respectively. The monostatic round-trip distance between BS-A and target $q$ is represented as
	\begin{equation}
		R_{\mathrm{A}}(\mathbf p_q)=2\|\mathbf p_q-\mathbf p_{\mathrm{A}}\|_2.
	\end{equation}
	For distributed BS $i$, the bistatic distance from BS-A to target $q$ and then to receiving BS $i$ is represented as
	\begin{equation}
		R_i(\mathbf p_q)=\|\mathbf p_q-\mathbf p_{\mathrm{A}}\|_2+\|\mathbf p_q-\mathbf p_i\|_2.
	\end{equation}
	The corresponding sample-domain delay is shown as
	\begin{equation}
		\tau_{i'}(\mathbf p_q)=\frac{R_{i'}(\mathbf p_q)f_{\rm{s}}}{c},
	\end{equation}
	where $f_{\rm{s}}$ is the sampling rate.
	Let $\omega(\mathbf p_q)$ denote the target bearing relative to the BS-A array. When target $q$ is the communication UAV, $\omega(\mathbf p_q)=\theta_{\rm{tx}}^{\rm{LoS}}$ under the same BS-A angular convention. Each sensing observation contains the superposed target echoes, the strong direct-path component, and receiver noise \cite{tao2025tvt}. 
	At each single-antenna distributed BS $i$, the target response is characterized primarily by the bistatic delay $\tau_i(\mathbf p_q)$. In contrast, the $N_{\rm{rx}}^{(\rm{A})}$-antenna BS-A preserves both the monostatic delay $\tau_{\mathrm{A}}(\mathbf p_q)$ and the array spatial response associated with $\omega(\mathbf p_q)$. Unless otherwise specified, the cooperative BSs are assumed to share a common timing reference \cite{delcourt2021tdoa,friedrich2021accurate}. 
	These delay and spatial signatures provide the measurement basis for the cooperative localization model developed in Section V.
	\vspace{-3mm}
	\section{Proposed UW-OCDM ISAC Waveform}
	To address the communication and sensing requirements formulated in Section II, this section introduces the OCDM modulation principles and develops the proposed UW-OCDM.
	
	\begin{figure}[!t]
		\captionsetup{font={footnotesize}, name = {Fig.}, labelsep = period}
		\centering
		\includegraphics[width=8.5cm]{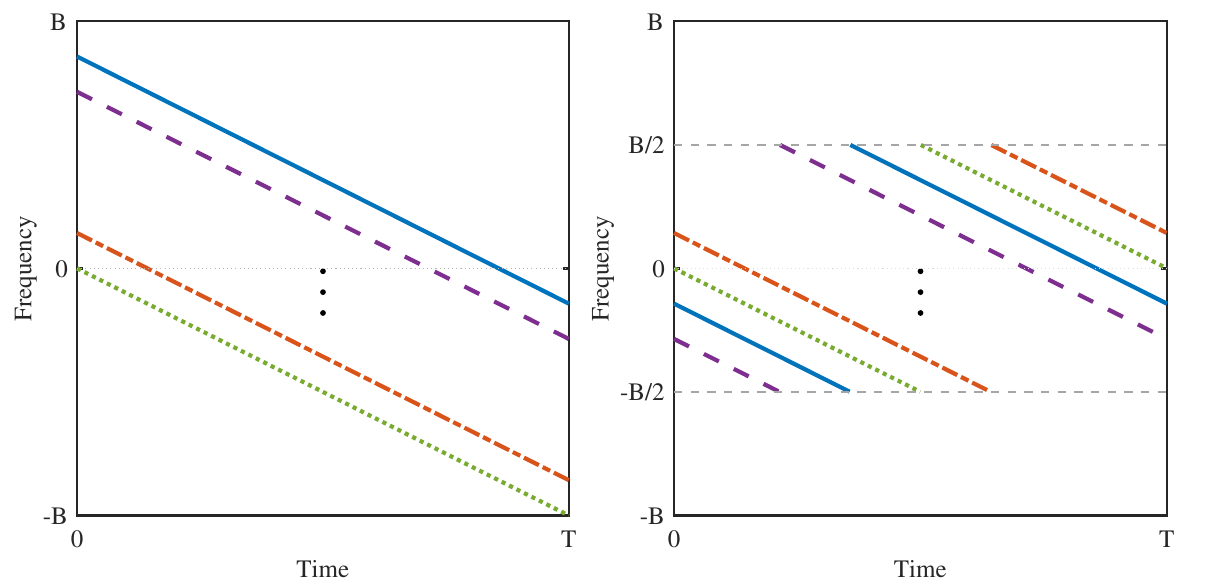}
		\caption{Time-frequency representation of the OCDM sub-chirps.}
		\label{Fig1}
		\vspace*{-5mm}
	\end{figure}
	
	\subsection{OCDM Preliminaries}
	OCDM is a multi-carrier modulation scheme based on the Fresnel transform \cite{ocdm}.
	Its core principle lies in utilizing a set of mutually orthogonal linear frequency modulation (chirp) signals to carry data \cite{ocdmisac,wan2026tcomm}.
	For an OCDM symbol with $N$ orthogonal chirp subcarriers (which can also be called sub-chirps), where $m$ denotes the sub-chirp index, the $m$-th time-domain analog sub-chirp signal $\psi_m(\tilde{t})$ is given by
	\begin{equation}
		\psi_{m}(\tilde{t})=\operatorname{rect}(\tilde{t} / T) e^{\jmath\frac{\pi}{4}} e^{-\jmath\pi\frac{N}{T^{2}}\left(\tilde{t} - \frac{m}{N}T \right)^{2}},
	\end{equation}
	where $\tilde{t}$ denotes the continuous-time variable and $T$ denotes the duration of an OCDM symbol.
	Owing to the staggered instantaneous frequencies of the individual sub-chirps, the total bandwidth of the synthesized analog OCDM signal, comprising $N$ continuous sub-chirps, spans from $-B$ to $B$, where $B = N/T$, as depicted in Fig.~\ref{Fig1}.
	Specifically, the lines with distinct colors and styles represent different orthogonal sub-chirps corresponding to distinct $m$.
	For practical digital implementations, the sampling rate is set to $f_{\rm{s}} = N/T$, with a discrete sampling interval of $T_{\rm{s}} = T/N$. 
	As illustrated in Fig.~\ref{Fig1}, due to the spectral folding effect \cite{ocdm}, the baseband sub-chirps are naturally confined to the interval $[-B/2, B/2)$ to satisfy bandwidth constraints.
	Consequently, the OCDM modulation can be formulated as a discrete-time process using DFnT, and the DFnT matrix $\boldsymbol{\Lambda} \in \mathbb{C}^{N \times N}$ is expressed as
	\begin{equation}
		\boldsymbol{\Lambda}[n,m] = \frac{1}{\sqrt{N}} e^{-\jmath\frac{\pi}{4}} \times 
		\begin{cases} 
			e^{\jmath\frac{\pi}{N}(n - m)^2}, & \langle N \rangle_2 = 0 \\ e^{\jmath\frac{\pi}{N}(n - m + \frac{1}{2})^2}, & \langle N \rangle_2 = 1 
		\end{cases},
	\end{equation}
	where $m,n\in\{0,\dots,N-1\}$.
	For simplicity, the following discussion focuses on the case where $\langle N \rangle_2 = 0$.
	
	Specifically, for the $t$-th transmit antenna, $N$ modulated symbols are first grouped to construct the Fresnel-domain vector $\mathbf{s}_t = [s_{t}[0], s_{t}[1], \dots, s_{t}[N-1]]^T \in \mathbb{C}^{N \times 1}$. 
	Through uniform discrete sampling, the discrete-time OCDM signal $\mathbf{x}_t$ is derived as
	\begin{equation}
		\begin{aligned}
			\mathbf{x}_t \left[n\right] &= \frac{1}{\sqrt{N}}\sum_{m=0}^{N-1} s_{t}[m] \psi_{m}\left(n\frac{T}{N}\right) \\
			&= \frac{1}{\sqrt{N}} e^{\jmath\frac{\pi}{4}} \sum_{m=0}^{N-1} s_{t}[m] e^{-\jmath\frac{\pi}{N}(n - m)^2},
		\end{aligned}
	\end{equation}
	which effectively aligns with the definition of the inverse discrete Fresnel transform (IDFnT) $\boldsymbol{\Lambda}^{H}$.
	Consequently, the discrete time-domain OCDM symbol $\mathbf{x}_t$ for the $t$-th transmit antenna can be formulated as
	\begin{equation}
		\mathbf{x}_t = \boldsymbol{\Lambda}^H \mathbf{s}_t.
	\end{equation}

\vspace{-3mm}	
\subsection{Proposed UW-OCDM Waveform Design} 

\begin{figure}[!t]
	\captionsetup{font={footnotesize}, name = {Fig.}, labelsep = period}
	\centering
	\includegraphics[width=9cm]{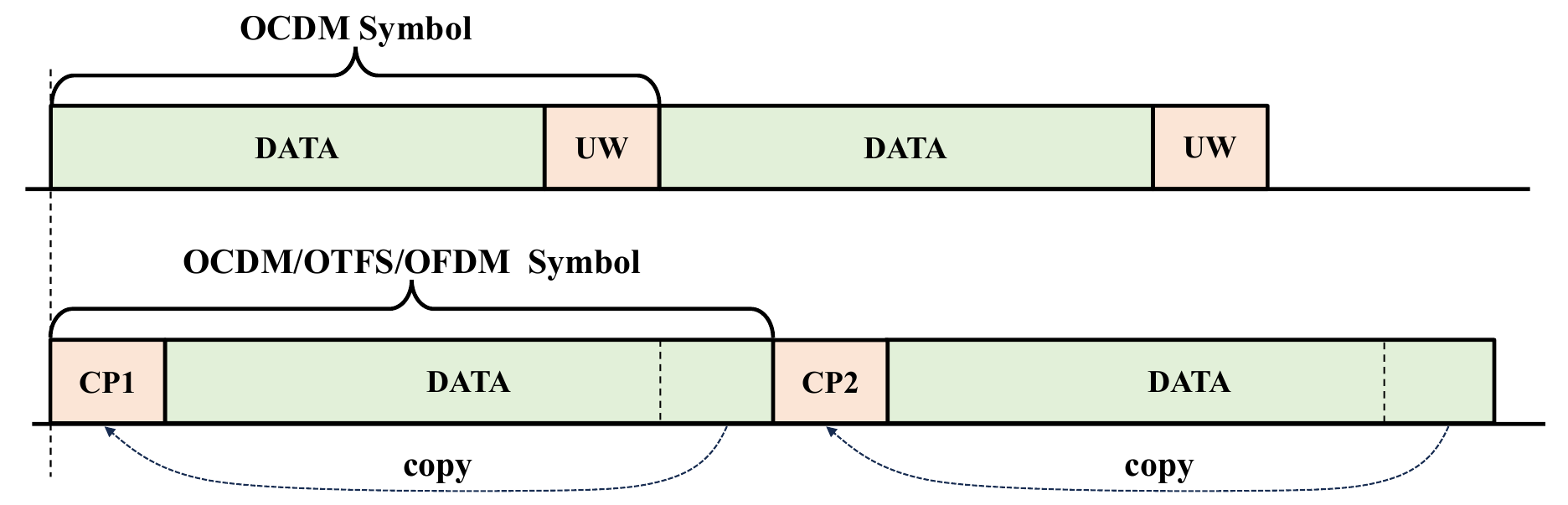} 
	\caption{Transmit data structure comparison: the proposed UW-OCDM and CP-OCDM, CP-OTFS, or CP-OFDM.}
	\label{fig:uw_cp_comparison}
	\vspace*{-5mm}
\end{figure}

\begin{figure}[!t]
	\captionsetup{font={footnotesize}, name = {Fig.}, labelsep = period}
	\centering
	\includegraphics[width=9cm]{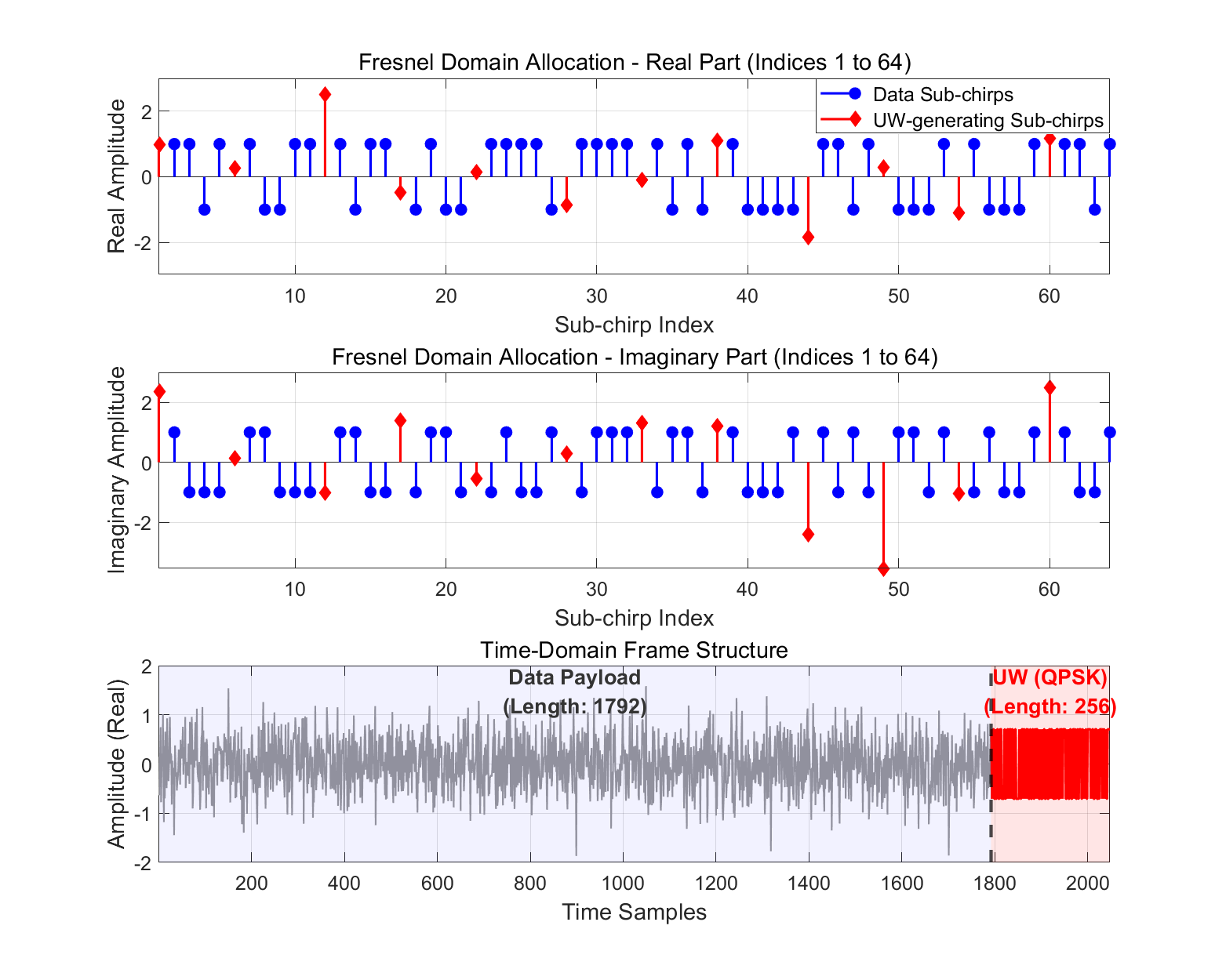}
	\caption{The proposed UW-OCDM symbol structure for $N=2048$, $N_{\rm{UW}}=256$, $N_{\rm{g}}=384$. UW is a unit-power quadrature phase shift keying-modulated sequence.}
	\label{fig:uw_structure}
	\vspace*{-6mm}
\end{figure}

To fulfill the dual requirements of communication and sensing in Section II, a novel UW-OCDM waveform design is proposed. 
As illustrated in Fig.~\ref{fig:uw_cp_comparison}, CP-OCDM mitigates ISI by prepending a data-dependent CP to the beginning of each symbol \cite{9346006}. 
In contrast, for the $t$-th transmit antenna, the tail of the discrete time-domain UW-OCDM symbol is replaced by a deterministic UW sequence $\mathbf{u}_t \in \mathbb{C}^{N_{\rm{UW}} \times 1}$, where $N_{\rm{UW}}$ is the time-domain UW sequence length. In this paper, a unit-power quadrature phase shift keying (QPSK) sequence\footnote{The UW is not restricted to QPSK. Other deterministic sequences may be used if known to both ends and compatible with the UW-generation constraint. QPSK is a practical constant-modulus choice providing reliable aperiodic timing and multi-antenna separation under the adopted receiver. Other sequences require joint design of correlation properties and antenna assignments.} is adopted as the UW.
The UW is embedded by reserving a subset of resources in the Fresnel domain.
The complete sub-chirp index set $\{0, \dots, N-1\}$ is partitioned into the data index set $\mathcal{I}_{\rm{d}}$ for allocating the data payload and the UW-generating index set $\mathcal{I}_{\rm{g}}$ reserved for UW construction.
The number of allocated UW-generating sub-chirps $N_{\rm{g}} \geq N_{\rm{UW}}$.
Consequently, the cardinalities of the corresponding index sets are defined as $|\mathcal{I}_{\rm{d}}| = N_{\rm{d}} = N - N_{\rm{g}}$ and $|\mathcal{I}_{\rm{g}}| = N_{\rm{g}}$.
As depicted in Fig.~\ref{fig:uw_structure}, these UW-generating sub-chirps are uniformly interleaved with the data sub-chirps within the Fresnel domain, which produces a distinct time-domain signal structure characterized by a random data payload followed by a deterministic UW.
Based on this subset partition, the signal vector $\mathbf{s}_t$ can be naturally decomposed into an effective data vector $\mathbf{d}_t = \mathbf{s}_t[\mathcal{I}_{\rm{d}}] \in \mathbb{C}^{N_{\rm{d}} \times 1}$ and an undetermined UW-generating vector $\mathbf{g}_t = \mathbf{s}_t[\mathcal{I}_{\rm{g}}] \in \mathbb{C}^{N_{\rm{g}} \times 1}$.

To satisfy the time-domain tail constraint of the UW, the last $N_{\rm{UW}}$ samples of $\mathbf{x}_t$ must exactly match $\mathbf{u}_t$. 
Therefore, let $\mathcal{I}_{\rm{tail}} = \{N - N_{\rm{UW}}, \dots, N-1\}$ denote the index set of the tail rows. We extract the corresponding rows from $\boldsymbol{\Lambda}^H$ to define the tail submatrix $\tilde{\boldsymbol{\Lambda}}_{\rm{tail}} = \boldsymbol{\Lambda}^H[\mathcal{I}_{\rm{tail}}, :] \in \mathbb{C}^{N_{\rm{UW}} \times N}$, which is further partitioned column-wise into a data block $\tilde{\boldsymbol{\Lambda}}_{\rm{tail,d}} = \tilde{\boldsymbol{\Lambda}}_{\rm{tail}}[:, \mathcal{I}_{\rm{d}}] \in \mathbb{C}^{N_{\rm{UW}} \times N_{\rm{d}}}$ and a UW-generating block $\tilde{\boldsymbol{\Lambda}}_{\rm{tail,g}} = \tilde{\boldsymbol{\Lambda}}_{\rm{tail}}[:, \mathcal{I}_{\rm{g}}] \in \mathbb{C}^{N_{\rm{UW}} \times N_{\rm{g}}}$. 
Accordingly, the UW constraint for the $t$-th antenna can be formulated as
\begin{align}
	\mathbf{u}_t = \tilde{\boldsymbol{\Lambda}}_{\rm{tail,d}} \mathbf{d}_t + \tilde{\boldsymbol{\Lambda}}_{\rm{tail,g}} \mathbf{g}_t.
\end{align}
By defining $\mathbf{b}_t= \mathbf{u}_t - \tilde{\boldsymbol{\Lambda}}_{\rm{tail,d}} \mathbf{d}_t$, the UW generation is transformed into solving $\tilde{\boldsymbol{\Lambda}}_{\rm{tail,g}} \mathbf{g}_t = \mathbf{b}_t$.
Given that $N_{\rm{g}} \geq N_{\rm{UW}}$, the matrix $\tilde{\boldsymbol{\Lambda}}_{\rm{tail,g}}$ has more columns than rows.
Consequently, the optimal UW-generating vector $\mathbf{g}_t^\star$ is determined by finding the minimum $\ell_2$-norm solution, which is efficiently computed via the right Moore-Penrose pseudoinverse, yielding the closed-form expression
\begin{align}
	\label{psedo}
\mathbf{g}_t^\star = \tilde{\boldsymbol{\Lambda}}_{\mathrm{tail,g}}^H \big(\tilde{\boldsymbol{\Lambda}}_{\mathrm{tail,g}} \tilde{\boldsymbol{\Lambda}}_{\mathrm{tail,g}}^H\big)^{-1} \mathbf{b}_t.
\end{align}
Finally, by mapping $\mathbf{g}_t^\star$ back to its corresponding UW-generating index positions and multiplexing it with the data vector $\mathbf{d}_t$, $\mathbf{x}_t$ that satisfies the UW constraint is generated.

\vspace{-3mm}
\subsection{Implementation Considerations}
\textbf{Computational Complexity:} The UW sequence $\mathbf{u}_t$ is pre-designed, while the pseudoinverse in \eqref{psedo} is only used to determine $\mathbf{g}_t^\star$. Since $\tilde{\boldsymbol{\Lambda}}_{\rm{tail,g}}$ depends only on $N$, $N_{\rm{UW}}$, $\mathcal{I}_{\rm{tail}}$, and $\mathcal{I}_{\rm{g}}$, rather than on the channel or UAV mobility, its pseudoinverse is computed once offline and stored at the transmitter with complexity $\mathcal{O}(N_{\rm{UW}}^{2}N_{\rm{g}}+N_{\rm{UW}}^{3})$. Online generation uses a fast IDFnT and the stored pseudoinverse with per-antenna and per-symbol complexity $\mathcal{O}(N\log N+N_{\rm{g}}N_{\rm{UW}})$, without any matrix inversion or pseudoinverse update.

\textbf{Resource Trade-off:} The choices of $N_{\rm{UW}}$ and $N_{\rm{g}}$ jointly determine the processing gain and communication efficiency. Increasing $N_{\rm{UW}}$ enlarges the ISI-free observation window, but may require a larger $N_{\rm{g}}$ to maintain feasible and well-conditioned UW synthesis. A larger $N_{\rm{g}}$ provides more synthesis freedom but leaves fewer data-bearing sub-chirps, thereby reducing net spectral efficiency.

\textbf{Peak-to-Average Power Ratio (PAPR) Considerations:} UW-OCDM is generated and transmitted by BS-A, while the UAV acts as the communication receiver and illuminated sensing target. Therefore, PAPR handling and power-amplifier (PA) linearization are performed in the BS-A transmit chain. OCDM generally exhibits PAPR characteristics comparable to those of OFDM \cite{ocdm}. Input power backoff and digital predistortion can mitigate PA nonlinear distortion in the BS-A transmit chain \cite{Yu2024ISACDPD}. Clipping can reduce waveform peaks, while filtering limits the resulting out-of-band spectral regrowth. The clipping level must be selected to limit distortion of the deterministic UW. The deterministic UW constraint can alter PAPR relative to an unconstrained data tail. PAPR-aware generator-matrix optimization developed for related UW waveforms offers a relevant future extension \cite{Rajabzadeh2021PAPR}.
                                                                
\vspace{-4mm}
\section{UW-Assisted Receiver Processing}
The UW is exploited at the receiver for timing synchronization and Doppler compensation.
Based on the aligned observations, this section further develops ST joint CE, low-complexity FDE, and data demodulation \cite{zhang2024network}.

\vspace{-3mm}
\subsection{UW-Assisted Timing Synchronization}
{Although the delay taps in Section II are indexed relative to the dominant LoS arrival, the absolute arrival position of each UW-OCDM symbol in the continuous received stream remains unknown. To distinguish the UW contributions during synchronization, the $N_{\rm{tx}}^{(\rm{A})}$ transmit antennas are assigned $\mathbf{u}_{t}$ with low pairwise aperiodic cross-correlation and low autocorrelation sidelobes.}
The sliding-correlation synchronizer assumes that the channel remains approximately quasi-static within one UW. Under the LoS-dominated GA channel with static scatterers, the path-wise Doppler magnitude is bounded by $f_{\rm{D},\max}=v_{\max}f_{\rm{c}}/c$, and the corresponding phase drift over $T_{\rm{UW}}=N_{\rm{UW}}T_{\rm{s}}$ is $\Delta\phi_{\max}=2\pi f_{\rm{D},\max}T_{\rm{UW}}$. For $f_{\rm{c}}=5$ GHz, $f_{\rm{s}}=40$ MHz, $N_{\rm{UW}}=256$, and $v_{\max}=150$ km/h, we obtain $T_{\rm{UW}}=6.4~\mu\mathrm{s}$, $f_{\rm{D},\max}\approx694.4$ Hz, and $\Delta\phi_{\max}\approx0.028$ rad. This small phase drift, together with the slowly varying path delays and amplitudes over this short interval, supports the quasi-static approximation used for UW correlation.

\begin{figure}[!t]
		\captionsetup{font={footnotesize}, name = {Fig.}, labelsep = period}
		\centering
		\includegraphics[width=6.5cm]{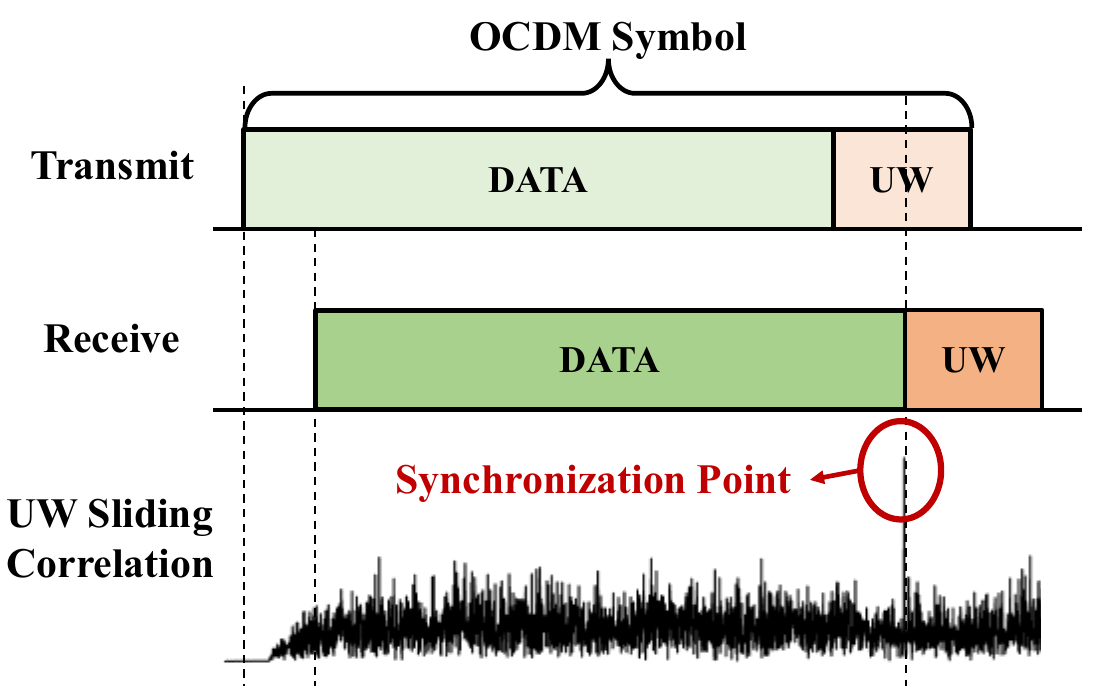} 
		\caption{Illustration of the UW-based sliding cross-correlation synchronization.}
		\label{fig:syn_logic}
		\vspace{-5mm}
\end{figure}

{As illustrated in Fig.~\ref{fig:syn_logic}, let $\ell_{\rm{syn}}$ denote a candidate starting index of the received UW window. The receiver evaluates the multi-antenna timing statistic as
\begin{equation}
	C[\ell_{\rm{syn}}]
	=
	\sum_{r=1}^{N_{\rm{rx}}^{(\rm{U})}}
	\sum_{t=1}^{N_{\rm{tx}}^{(\rm{A})}}
	\left|
		\sum_{g'=0}^{N_{\rm{UW}}-1}
		y_r[\ell_{\rm{syn}}+g']u_t^*[g']
	\right|^2,
\end{equation}
which noncoherently accumulates the correlation energies across receive antennas and UWs.
The strongest valid peak in $C[\ell_{\rm{syn}}]$ is associated with the dominant-path UW arrival.
Its sample-level starting index is estimated as
\begin{equation}
	{\hat\ell_{\rm{UW}}}
	=
	\arg\max_{\ell_{\rm{syn}}} C[\ell_{\rm{syn}}].
\end{equation}
The timing decision is accepted only when $C[{\hat\ell_{\rm{UW}}}]\geq\gamma_{\rm{syn}}$. Therefore, the estimated starting index of the UW-OCDM symbol is represented as
	\begin{equation}
	{\hat\ell_{\rm{start}}}
	=
	{\hat\ell_{\rm{UW}}}-(N-N_{\rm{UW}}).
	\end{equation}}
In Section II, $n_{\rm{abs}}$ denotes the absolute sample index. After alignment, $b\in\{0,1,\dots\}$ indexes the aligned UW-OCDM symbols, with $n_{\rm{abs}}(b,n)\triangleq\hat\ell_{\rm{start}}+bN+n$ denoting the corresponding absolute index. The aligned samples are thus given by $y_{r,b}[n]\triangleq y_r[n_{\rm{abs}}(b,n)]$, where consecutive UW observations are extracted for Doppler estimation.

\vspace{-5mm}
\vspace{0pt plus -1.5ex}
\subsection{UW-Assisted Doppler Mitigation}
\vspace{0pt plus -0.5ex}
\vspace{-1mm}
\begin{figure}[!t]
\captionsetup{font={footnotesize}, name = {Fig.}, labelsep = period}
\centering
\includegraphics[width=8.5cm]{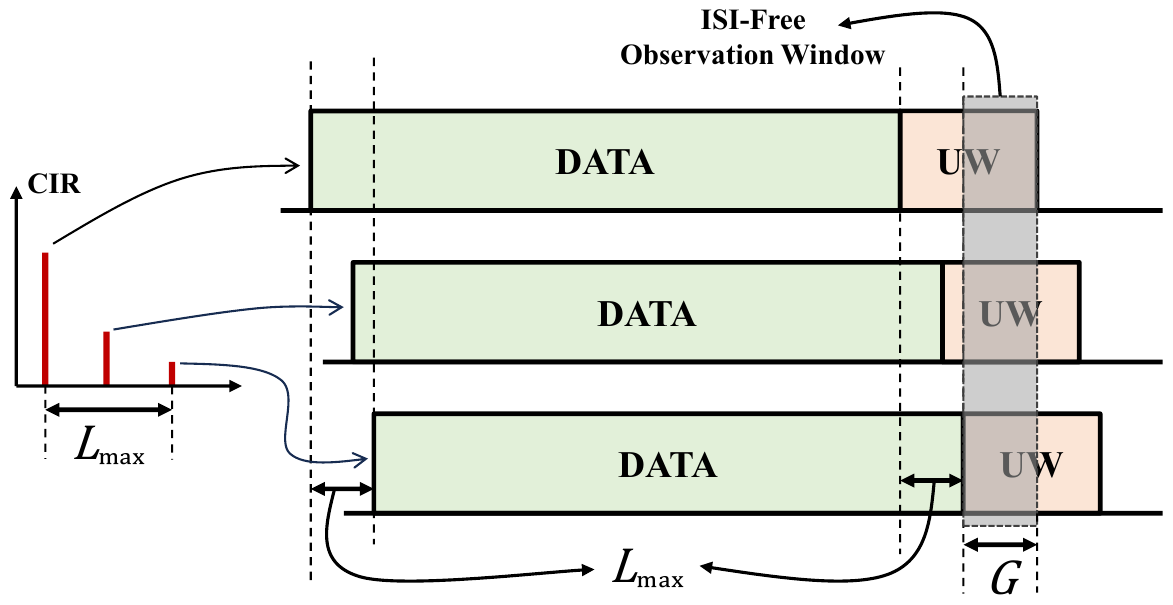} 
\caption{Illustration of the multipath delay spread effect and the construction of the ISI-free observation window. The maximum delay spread $L_{\max}$ dictates the discarded portion of the UW to avoid data-induced ISI.}
\label{fig:obs_window}
\vspace{-5mm}
\end{figure}
Following timing synchronization, the receiver selects $X_{\rm{num}}$ UWs indexed by $\chi\in\{0,\dots,X_{\rm{num}}-1\}$, separated by $D$ symbol intervals. Taking the first aligned symbol as the observation origin, $b_\chi\triangleq\chi D$ denotes the index of the aligned UW-OCDM symbol containing the $\chi$-th selected UW. As shown in Fig.~\ref{fig:obs_window}, the first $L_{\max}$ UW samples affected by data-induced ISI are discarded, leaving $G=N_{\rm{UW}}-L_{\max}$ clean samples for each UW.
Let $\mathbf y_{r,\chi}^{\rm{UW}}\triangleq[y_{r,b_\chi}[N-N_{\rm{UW}}],\dots,y_{r,b_\chi}[N-1]]^T$ be the $N_{\rm{UW}}$-sample UW observation. The ISI-free segment is $\bar{\mathbf u}_{r,\chi}\triangleq\mathcal{T}_G\{\mathbf y_{r,\chi}^{\rm{UW}}\}\in\mathbb C^{G\times1}$, where $\mathcal{T}_G\{\cdot\}$ retains its last $G$ entries and $g\in\{0,\dots,G-1\}$ indexes the resulting samples.

Since each BS-A transmit antenna repeats its assigned UW and the propagation response remains approximately unchanged over the span $(X_{\rm{num}}-1)DNT_{\rm{s}}$, corresponding samples in adjacent observations differ primarily by the common LoS Doppler-induced phase rotation.
{
Therefore, the pairwise Doppler-induced phase rotation can be formulated as
\begin{equation}
		\bar{u}_{r,\chi+1}[g] \approx \bar{u}_{r,\chi}[g] e^{\jmath \varphi} + \nu_{r,\chi}[g],
\end{equation}
where $\varphi\triangleq 2\pi f_{\rm{D,LoS}} D N T_{\rm{s}}$ denotes the accumulated phase difference between two adjacent UW observations, and $\nu_{r,\chi}[g]$ denotes the effective pairwise disturbance. Under the adopted independent and identically distributed complex Gaussian approximation, $\nu_{r,\chi}[g]\sim\mathcal{CN}(0,\sigma_\nu^2)$, while $\varphi$ is common to all antennas, UW pairs, and samples. Under this conditional pairwise model, the maximum-likelihood (ML) estimation of $\varphi$ is obtained by minimizing the negative log-likelihood over all observations, which is equivalently
\begin{align}
	\hat{\varphi}_{\rm{ML}}
	&=
	\arg\min_{\varphi}
	\sum_{\chi=0}^{X_{\rm{num}}-2}
	\sum_{r=1}^{N_{\rm{rx}}^{(\rm{U})}}
	\sum_{g=0}^{G-1}
	\left|
	\bar{u}_{r,\chi+1}[g]
	-
	\bar{u}_{r,\chi}[g]e^{\jmath\varphi}
	\right|^2
	\nonumber\\
	&=
	\arg\max_{\varphi}
	\operatorname{Re}
	\left\{
	\sum_{\chi=0}^{X_{\rm{num}}-2}
	\sum_{r=1}^{N_{\rm{rx}}^{(\rm{U})}}
	\sum_{g=0}^{G-1}
	\bar{u}_{r,\chi+1}[g]
	\bar{u}_{r,\chi}^{*}[g]
	e^{-\jmath\varphi}
	\right\}.
	\label{eq:ml_phi}
\end{align}
The global cross-correlation statistic is defined as
	\begin{equation}
		\label{rho_global}
		\rho_{\rm{global}} = \sum_{\chi=0}^{X_{\rm{num}}-2} \sum_{r=1}^{N_{\rm{rx}}^{(\rm{U})}} \sum_{g=0}^{G-1} \bar{u}_{r,\chi+1}[g] \left( \bar{u}_{r,\chi}[g] \right)^*.
	\end{equation}
With this coherently accumulated correlation statistic, the problem reduces to
\begin{equation}
	\hat{\varphi}_{\rm{ML}}
	=
	\arg\max_{\varphi}
	\operatorname{Re}
	\left\{
	\rho_{\rm{global}}e^{-\jmath\varphi}
	\right\}.
\end{equation}
Writing $\rho_{\rm{global}}=|\rho_{\rm{global}}|e^{\jmath\angle\rho_{\rm{global}}}$, we can obtain $\operatorname{Re}\{\rho_{\rm{global}}e^{-\jmath\varphi}\}=|\rho_{\rm{global}}|\cos(\angle\rho_{\rm{global}}-\varphi)$, which is maximized when $\varphi=\angle\rho_{\rm{global}}$ (mod $2\pi$). Hence, under the adopted model, $\angle\rho_{\rm{global}}$ is the maximum-likelihood estimate of the accumulated phase difference, and the Doppler frequency offset is estimated as
	\begin{equation}
		\hat{f}_{\rm{D}} = \frac{\angle \rho_{\rm{global}}}{2\pi D N T_{\rm{s}}}.
	\end{equation}
}
It is worth noting that while increasing $X_{\rm{num}}$ enhances noise suppression through more accumulation, the maximum Doppler estimation range is governed solely by the single-step interval. Since $\angle(\cdot)$ is defined modulo $2\pi$, the accumulated phase must satisfy $2\pi f_{\rm{D,\max}} D N T_{\rm{s}} < \pi$ to prevent phase ambiguity.
	
Once $\hat{f}_{\rm{D}}$ is obtained, the receiver de-rotates each symbol using the relative sample position $bN+n=n_{\rm{abs}}(b,n)-\hat\ell_{\rm{start}}$. The compensated received signal is formulated as
\begin{equation}
	\tilde{y}_{r,b}[n] = y_{r,b}[n] e^{-\jmath 2\pi \hat{f}_{\rm{D}} (bN+n) T_{\rm{s}}}.
\end{equation}
Similarly, let $\tilde{\mathbf y}_{r,\chi}^{\rm{UW}}\triangleq[\tilde y_{r,b_\chi}[N-N_{\rm{UW}}],\dots,\tilde y_{r,b_\chi}[N-1]]^T$. The corresponding ISI-free segment $\tilde{\mathbf u}_{r,\chi}\triangleq\mathcal{T}_G\{\tilde{\mathbf y}_{r,\chi}^{\rm{UW}}\}$ is passed to the subsequent ST joint CE.

\vspace{-5mm}
\subsection{MMV-OMP-Based Joint Channel Estimation}
In the proposed MIMO UW-OCDM, the GA multipath channel exhibits inherent sparsity \cite{9898900}. 
	Furthermore, the physical propagation environment dictates that the multipath delays remain virtually invariant across the receive antennas \cite{ke2020tsp} and consecutive UW-OCDM symbols, even though their complex path gains may vary. 
	To fully exploit this joint ST sparsity, we formulate the CE as an MMV CS problem \cite{li2023thzris}.
Specifically, to fully exploit the correlation across the entire observation duration, we construct a global UW observation $\mathbf{U} \in \mathbb{C}^{G \times N_{\rm{rx}}^{(\rm{U})}X_{\rm{num}}}$ by concatenating all the Doppler-compensated ISI-free UW segments $\tilde{\mathbf{u}}_{r,\chi}$ for all $N_{\rm{rx}}^{(\rm{U})}$ antennas, represented as
	\begin{equation}
		\mathbf{U} = \left[ \tilde{\mathbf{u}}_{1,0}, \dots, \tilde{\mathbf{u}}_{1,X_{\rm{num}}-1}, \dots, \tilde{\mathbf{u}}_{N_{\rm{rx}}^{(\rm{U})},0}, \dots, \tilde{\mathbf{u}}_{N_{\rm{rx}}^{(\rm{U})},X_{\rm{num}}-1} \right].
\end{equation}
The reference UW matrix $\boldsymbol{\Psi} \in \mathbb{C}^{G \times N_{\rm{tx}}^{(\rm{A})}L_{\max}}$ is constructed by horizontally concatenating the antenna-specific UW convolution submatrices as $\boldsymbol{\Psi}=[\boldsymbol{\Psi}_1,\dots,\boldsymbol{\Psi}_{N_{\rm{tx}}^{(\rm{A})}}]$, where $\boldsymbol{\Psi}_t\in\mathbb C^{G\times L_{\max}}$ is Toeplitz and its $(g,l)$-th element is $[\boldsymbol{\Psi}_t]_{g,l}=u_t[L_{\max}+g-l]$. Accordingly, column $(t-1)L_{\max}+l$ of $\boldsymbol{\Psi}$ corresponds to delay tap $l$ of transmit antenna $t$.
Therefore, the MMV CS problem can be expressed as
\begin{align}
	\mathbf{U} = \boldsymbol{\Psi} \mathbf{H}_{\rm{all}} + \mathbf{W},
\end{align}
where $\mathbf{H}_{\rm{all}} \in \mathbb{C}^{N_{\rm{tx}}^{(\rm{A})}L_{\max} \times N_{\rm{rx}}^{(\rm{U})}X_{\rm{num}}}$ is the row-sparse channel coefficient matrix to be estimated, and $\mathbf{W}$ collects effective noise and residual modeling errors. 
To relate the CE model to the CIR in Section II, define the Doppler-compensated CIR as $\tilde h_{r,t,b}[n,l]\triangleq h_{r,t}[n_{\rm{abs}}(b,n),l]e^{-\jmath2\pi\hat f_{\rm{D}}(bN+n)T_{\rm{s}}}$. The phase referenced to $\hat\ell_{\rm{start}}$ is constant over the aligned symbols and is absorbed into the estimated CIR. Over the ISI-free UW window, $\tilde h_{r,t,b_\chi}[N-N_{\rm{UW}}+L_{\max}+g,l]\approx h_{r,t,\chi}[l]$. Accordingly, $\mathbf{h}_{r,t,\chi} = \big[h_{r,t,\chi}[0], \dots, h_{r,t,\chi}[L_{\max}-1]\big]^T \in \mathbb{C}^{L_{\max} \times 1}$ denotes the local CIR snapshot.
By concatenating $\mathbf{h}_{r,t,\chi}$, the composite vector $\mathbf{h}_{r,\chi} = \big[\mathbf{h}_{r,1,\chi}^T, \mathbf{h}_{r,2,\chi}^T, \dots, \mathbf{h}_{r,N_{\rm{tx}}^{(\rm{A})},\chi}^T \big]^T \in \mathbb{C}^{N_{\rm{tx}}^{(\rm{A})}L_{\max} \times 1}$ is obtained.
Consequently, $\mathbf{H}_{\rm{all}}$ can be represented as $\mathbf{H}_{\rm{all}} = \left[ \mathbf{h}_{1,0}, \dots, \mathbf{h}_{1,X_{\rm{num}}-1}, \dots, \mathbf{h}_{N_{\rm{rx}}^{(\rm{U})},0}, \dots, \mathbf{h}_{N_{\rm{rx}}^{(\rm{U})},X_{\rm{num}}-1} \right]$.
Although the complex path gains may vary across antennas and UWs, the physical path delays remain approximately common over the considered observation duration. Hence, the columns of $\mathbf{H}_{\rm{all}}$ share the same nonzero row support, exhibiting joint row-sparsity.
To resolve this problem, we employ MMV-OMP to identify the joint support set and subsequently obtain the estimate $\hat{\mathbf{H}}_{\rm{all}}$. 

Finally, for each adjacent pair, the estimates $\hat{\mathbf{h}}_{r,\chi}$ and $\hat{\mathbf{h}}_{r,\chi+1}$ are phase-aligned to form a CIR estimate as
\begin{align}
	\Delta \boldsymbol{\theta} &= \angle \left( \big(\hat{\mathbf{h}}_{r,\chi}\big)^* \odot \hat{\mathbf{h}}_{r,\chi+1} \right),  \\
	\hat{\mathbf{h}}_{r,\chi}^{\rm{avg}} &= \frac{1}{2} \left( \big|\hat{\mathbf{h}}_{r,\chi}\big| + \big|\hat{\mathbf{h}}_{r,\chi+1}\big| \right) \odot e^{  \jmath \left( \angle \hat{\mathbf{h}}_{r,\chi} + \frac{\Delta \boldsymbol{\theta}}{2} \right) }.
\end{align}

\vspace{-2mm}
\subsection{Frequency-Domain Equalization and Data Demodulation}
{
		For the $\chi$-th UW pair, define $\mathcal D_\chi\triangleq\{b_\chi+1,\dots,b_{\chi+1}\}$. 
		Under the quasi-static approximation, for a fixed $b\in\mathcal D_\chi$, let 
		$\hat{\mathbf{h}}_{r,\chi}^{\rm{avg}}
		=
		\big[(\hat{\mathbf{h}}^{\rm{avg}}_{r,1,\chi})^T,\dots,
		(\hat{\mathbf{h}}^{\rm{avg}}_{r,N_{\rm{tx}}^{(\rm{A})},\chi})^T \big]^T
		\in \mathbb{C}^{N_{\rm{tx}}^{(\rm{A})}L_{\max} \times 1}$,
		where $\hat{\mathbf h}^{\rm{avg}}_{r,t,\chi}$ denotes the representative CIR for the $t$-th transmit antenna obtained from the $\chi$-th UW pair. 
		For notational simplicity, $b$ and $\chi$ are omitted below when unambiguous. 
		The receiver then performs FDE and data demodulation.}
In a continuous transmission stream, for each transmit antenna, the tail of every UW-OCDM symbol is constrained to the same UW. 
Provided that $N_{\rm{UW}}\geq L_{\max}$, the ISI originating from the preceding symbol's UW is equivalent to the current symbol's own tail UW cyclically wrapping around to its leading edge. 
	Under the post-compensation symbol-wise quasi-static approximation, this CP-equivalent property allows the time-domain linear convolution to be represented as an $N$-point circular convolution, which enables low-complexity single-tap FDE \cite{UW-OFDM2}\footnote{The low-complexity here refers to the equalization stage. The inverse DFT and DFnT retain $\mathcal{O}(N\log N)$ transform complexity. The total receiver complexity is nevertheless higher than that of least-squares (LS)-based CP-OFDM mainly because MMV-OMP CE requires iterative dictionary correlations and support-restricted LS updates to exploit the common sparse support across antennas and UWs for improved CE accuracy.}.
Therefore, the equivalent frequency-domain MIMO channel matrix $\hat{\mathbf{H}}_{\mathrm{freq}}[m] \in \mathbb{C}^{N_{\rm{rx}}^{(\rm{U})} \times N_{\rm{tx}}^{(\rm{A})}}$ for the $m$-th frequency bin is constructed, with its $(r, t)$-th element given by
\begin{equation}
	\hat{h}^{\mathrm{freq}}_{r,t}[m] = \sum_{l=0}^{L_{\max}-1} \hat{h}^{\mathrm{avg}}_{r,t}[l] \cdot e^{-\jmath \frac{2\pi}{N} m l}.
\end{equation}
Concurrently, the compensated samples of the current aligned symbol are stacked as $\tilde{\mathbf y}_{r,b}\triangleq[\tilde y_{r,b}[0],\dots,\tilde y_{r,b}[N-1]]^T\in\mathbb C^{N\times1}$. For brevity, let $\tilde{\mathbf y}_r\triangleq\tilde{\mathbf y}_{r,b}$. It is converted via $\mathbf{y}^{\mathrm{freq}}_r = \mathbf{F}_N \tilde{\mathbf{y}}_r$. 
By extracting the $m$-th frequency-bin components across all $N_{\rm{rx}}^{(\rm{U})}$ antennas, the observation vector is formulated as $\mathbf{y}_{\mathrm{freq}}[m] = \big[y^{\mathrm{freq}}_1[m], y^{\mathrm{freq}}_2[m], \dots, y^{\mathrm{freq}}_{N_{\rm{rx}}^{(\rm{U})}}[m] \big]^T \in \mathbb{C}^{N_{\rm{rx}}^{(\rm{U})} \times 1}$.
In this paper, the minimum mean square error (MMSE) equalizer is employed.
The equalization weight matrix $\mathbf{W}_{\rm{MMSE}}[m] \in \mathbb{C}^{N_{\rm{tx}}^{(\rm{A})} \times N_{\rm{rx}}^{(\rm{U})}}$ for the $m$-th frequency bin is represented as
\begin{align}
	\mathbf{W}_{\rm{MMSE}}[m] = \left( \hat{\mathbf{H}}_{\mathrm{freq}}^{H}[m] \hat{\mathbf{H}}_{\mathrm{freq}}[m] + \sigma_w^2 \mathbf{I}_{N_{\rm{tx}}^{(\rm{A})}} \right)^{-1} \hat{\mathbf{H}}_{\mathrm{freq}}^H[m],
\end{align}
where $\sigma_w^2$ is the noise variance.
Consequently, the frequency-domain signal is then equalized by
\begin{align}
	\hat{\mathbf{x}}_{\rm{freq}}[m] = \mathbf{W}_{\rm{MMSE}}[m] \mathbf{y}_{\rm{freq}}[m],
\end{align}
where $\hat{\mathbf{x}}_{\rm{freq}}[m] \in \mathbb{C}^{N_{\rm{tx}}^{(\rm{A})} \times 1}$ represents the equalized vector for all transmit antennas at the $m$-th frequency bin.

For the $t$-th transmit antenna, the receiver extracts the corresponding element from $\hat{\mathbf{x}}_{\rm{freq}}[m]$ across each frequency bin and concatenates them vertically in frequency order.
Therefore, $\hat{\mathbf{x}}^{\rm{freq}}_{t} \in \mathbb{C}^{N \times 1}$ can be obtained, which is then transformed back to the time-domain via an $N$-point inverse DFT and subsequently decoupled by the DFnT matrix $\boldsymbol{\Lambda}$ to recover the Fresnel-domain symbol vector $\hat{\mathbf{s}}_t \in \mathbb{C}^{N \times 1}$, formulated as
\begin{equation}
	\hat{\mathbf{s}}_t = \boldsymbol{\Lambda} \mathbf{F}_N^H \hat{\mathbf{x}}_{t}^{\rm{freq}}.
\end{equation}
Finally, data extraction and demodulation are performed. According to the proposed resource partition strategy, the pure data payload vector is extracted from the data index set as $\hat{\mathbf{d}}_t = \hat{\mathbf{s}}_t[\mathcal{I}_{\rm{d}}]$. Standard quadrature amplitude modulation (QAM) demodulation is then applied to $\hat{\mathbf{d}}_t$ to recover the transmit binary bit stream.

\vspace{-3mm}
\section{Multi-Static UAV Sensing and Localization}
Building on the multi-static sensing model in Section II-B, we construct a UW-based cooperative localization framework using the delay signatures at the distributed BSs and the ST signature at BS-A \cite{gao2025jsac_hfbs}. The processing pipeline comprises local sensing-dictionary reconstruction and DPI suppression, followed by hierarchical target extraction and SIC.

\vspace{-4mm}
\subsection{Sensing Dictionary and Direct-Path Suppression}
	For each sensing BS $i'\in\mathcal I$, let $\tilde{\mathbf u}_{i'}^{\rm{obs}}\in\mathbb C^{N_{i'}\times1}$ denote its local UW sensing observation extracted relative to the common BS-A transmit-block timing using the $G$-sample ISI-free window, where $N_{\mathrm{A}}=GN_{\rm{rx}}^{(\rm{A})}$ and $N_i=G$. In the dictionary, $\mathbf p$ denotes a candidate position. Each $\mathbf p$ determines $\tau_{i'}(\mathbf p)$ and $\omega(\mathbf p)$ through the geometry defined in Section II-B. Assuming array spacing $d_{\rm{A}}$, the effective transmit UW toward $\mathbf p$ is represented as
	\begin{equation}
		{\mathbf{u}_{\rm{comb}}(\mathbf{p}) = \sum_{t=1}^{N_{\rm{tx}}^{(\rm{A})}} \mathbf{u}_t e^{-\jmath \frac{2\pi d_{\rm{A}}}{\lambda} (t-1) \cos \omega(\mathbf{p})}.}
	\end{equation}		
	By exploiting the Fourier time-shifting property, the delay atom at BS $i'$ is constructed as $\boldsymbol{\phi}_{i'}(\mathbf p)\in\mathbb C^{G\times1}$, which is given by		
	\begin{equation}		
		\boldsymbol{\phi}_{i'}(\mathbf p) = \mathcal{T}_G \Big\{ \mathbf{F}_{N_{\rm{UW}}}^{-1} \big( (\mathbf{F}_{N_{\rm{UW}}} \mathbf{u}_{\rm{comb}}(\mathbf{p})) \odot \mathbf{e}_{i'}(\mathbf p) \big) \Big\},
	\end{equation}
	where $\mathbf{e}_{i'}(\mathbf p) \in \mathbb{C}^{N_{\rm{UW}} \times 1}$ applies the delay $\tau_{i'}(\mathbf p)$ to each frequency component, with its $j$-th element defined as
\begin{equation}
	\mathbf{e}_{i'}(\mathbf p)[j] = \exp \left( -\jmath \frac{2\pi}{N_{\rm{UW}}} \kappa(j) \cdot \tau_{i'}(\mathbf{p}) \right),
\end{equation}
where $\kappa(j)$ maps the standard DFT index to the zero-centered physical frequency, given by $\kappa(j) = j$ for $0 \le j < N_{\rm{UW}}/2$, and $\kappa(j) = j - N_{\rm{UW}}$ otherwise.
For each distributed BS $i$, $\boldsymbol{\phi}_i(\mathbf p)$ is its complete local atom. For BS-A, the full ST atom is $\boldsymbol{\phi}_{\mathrm{A,ST}}(\mathbf p)=\mathbf a_{\rm{rx}}^{(\rm{A})}(\omega(\mathbf p))\otimes\boldsymbol{\phi}_{\mathrm{A}}(\mathbf p)$, where $\mathbf a_{\rm{rx}}^{(\rm{A})}(\cdot)\in\mathbb C^{N_{\rm{rx}}^{(\rm{A})}\times1}$ denotes the receive steering vector of BS-A.
Because the deterministic UW and system configuration are pre-stored, each cooperative BS can reconstruct its local sensing atoms without acquiring per-symbol payload data. In contrast, a random-payload-based coherent sensing design requires the instantaneous transmit waveform to be obtained through real-time sharing or local decoding, incurring transmit-reference sharing overhead or a decoding burden, whereas introducing an additional deterministic reference instead reduces spectral and ISAC efficiency.

However, in multi-static cooperative systems, receiving BSs typically lie within the LoS of BS-A. 
The resulting direct path (DP) energy can be tens of decibels (dB) stronger than the scattered UAV echoes, severely monopolizing the receiver's dynamic range. Therefore, DPI suppression is essential. 
For each BS $i'$, the DP atom $\boldsymbol{\phi}_{i'}^{\rm{DP}}$ is constructed from the known station geometry or a calibrated direct channel. At BS-A, it represents the calibrated self-interference ST signature. At distributed BS $i$, the DP atom is constructed using the sample-domain propagation delay $\tau_i^{\rm{DP}}=\|\mathbf p_i-\mathbf p_{\rm{A}}\|_2f_{\rm{s}}/c$ and the transmit-array phase determined by the known bearing from BS-A to BS $i$.
The DP component is independently canceled at each BS $i'$ by projecting $\tilde{\mathbf{u}}^{\rm{obs}}_{i'}$ onto the orthogonal complement subspace of $\boldsymbol{\phi}^{\mathrm{DP}}_{i'}$. The cleaned residual signal $\tilde{\mathbf{u}}^{\rm{res}}_{i'}$ is thus obtained as
\begin{equation}
	\label{u_res_local}
	\tilde{\mathbf{u}}^{\rm{res}}_{i'} = \mathbf{P}_{i'}^{\perp}\tilde{\mathbf{u}}^{\rm{obs}}_{i'}= \left( \mathbf{I}_{N_{i'}} - \frac{\boldsymbol{\phi}^{\mathrm{DP}}_{i'} (\boldsymbol{\phi}^{\mathrm{DP}}_{i'})^H}{\|\boldsymbol{\phi}^{\mathrm{DP}}_{i'}\|_2^2} \right) \tilde{\mathbf{u}}^{\rm{obs}}_{i'}.
\end{equation}
Then, the system utilizes $\{ \tilde{\mathbf{u}}^{\rm{res}}_{i'} \}_{i'\in\mathcal I}$ as the inputs to sequentially execute the multi-target searches.
To prevent subspace mismatch in subsequent processing, all ST and delay atoms generated hereafter must also be projected onto their respective local orthogonal subspaces, which is represented as
	\begin{equation}
		\tilde{\boldsymbol{\phi}}_i(\mathbf{p}) = \mathbf{P}_i^{\perp} \boldsymbol{\phi}_i(\mathbf{p}).
	\end{equation}
	\begin{equation}
		\tilde{\boldsymbol{\phi}}_{\mathrm{A,ST}}(\mathbf{p}) = \mathbf{P}_{\mathrm{A}}^{\perp} \boldsymbol{\phi}_{\mathrm{A,ST}}(\mathbf{p}).
	\end{equation}

\vspace{-5mm}
\subsection{Iterative Coarse-to-Fine Multi-Target Localization}
Since the total number of low-altitude UAV targets is typically unknown and much smaller than the number of spatial grids, the cooperative localization can be transformed into a dynamic sparse recovery problem \cite{UAVCS}.
A hierarchical algorithm performs at most $K_{\max}$ extraction attempts, where $k \in \{1,\dots,K_{\max}\}$ indexes the processing attempt. At BS $i'$, the local residual and active dictionary of accepted projected atoms are initialized as $\mathbf r_{i'}^{(0)}=\tilde{\mathbf u}_{i'}^{\rm{res}}$ and $\widetilde{\boldsymbol\Phi}_{i'}^{(0)}=\emptyset$, respectively. If an attempt is rejected, both states are carried forward unchanged.

\subsubsection{Kinematic Prediction and Target Prioritization}
Before initiating the iteration, the system prioritizes actively tracked targets from previous symbols.
For each established track $q\in\{1,\dots,Q_{\rm{trk}}\}$, its expected position is extrapolated as
\begin{equation}
	\label{ppred}
	\mathbf{p}_{\rm{pred}, q} = \mathbf{p}_{\rm{last}, q} + \mathbf{v}_{\rm{est}, q} \Delta T,
\end{equation}
where $Q_{\rm{trk}}$ is the number of existing tracks, $\mathbf p_{\rm{last},q}$ and $\mathbf v_{\rm{est},q}$ are their previous position and velocity estimates, and $\Delta T$ is the tracking interval.
The tracks are prioritized using the joint normalized matching score, represented as
\begin{equation}
	\label{eq}
	E_q = \big| \bar{\boldsymbol{\phi}}_{\mathrm{A,ST}}^H(\mathbf{p}_{\rm{pred}, q}) \bar{\mathbf{r}}_{\mathrm{A}}^{(0)} \big| + \sum_{i=1}^I \big| \bar{\boldsymbol{\phi}}_i^H(\mathbf{p}_{\rm{pred}, q}) \bar{\mathbf{r}}_i^{(0)} \big|,
\end{equation}
where \textbf{an overbar denotes $\ell_2$ normalization in this section}. Sorting $\{E_q\}$ in descending order yields $\{\eta_k\}_{k=1}^{K_{\max}}$, where $\eta_k$ is the fixed scheduled-track index during attempt $k$, while $\eta_k=0$ denotes a global-acquisition attempt after all existing tracks have been scheduled. This ordering prioritizes strong tracked targets before weak ones.

\subsubsection{Prediction-Aided Coarse Candidate Acquisition}
Candidate centers are acquired through two paths at attempt $k$.
First, if $\eta_k\neq 0$, acquisition uses the predicted position as the candidate center.
Denoting this center by $\mathbf p_{\rm{c},1}^{(k)}$, we have $\mathbf p_{\rm{c},1}^{(k)}=\mathbf p_{\rm{pred},\eta_k}$ and $\mathcal C^{(k)}=\{\mathbf p_{\rm{c},1}^{(k)}\}$, which directly enters local refinement without global coarse search. If this predicted candidate fails verification, global acquisition is invoked once as a fallback.
Second, if $\eta_k=0$ or the fallback is triggered, candidate centers are acquired from the global coarse grid.

To avoid repeatedly detecting accepted targets, subsequent global searches exclude their neighboring coarse grids. Let $\mathcal P_{\rm{coa}}=\{\mathbf p_\nu\}_{\nu=1}^{N_{\rm{c}}}$ be the $N_{\rm{c}}$-point coarse 3D grid, and define the radius-$d$ neighborhood centered at $\mathbf p_0$ as $\mathcal B(\mathbf p_0,d)=\{\mathbf p\in\mathbb R^3:\|\mathbf p-\mathbf p_0\|_2\leq d\}$. Let $\widehat{\mathcal P}^{(k-1)}$ contain the positions accepted in preceding attempts, with $\widehat{\mathcal P}^{(0)}=\emptyset$, and let $d_{\rm{NMS}}>0$ denote the spatial suppression radius. Accordingly, global acquisition searches the available grid $\mathcal P_{\rm{coa}}^{(k)}=\mathcal P_{\rm{coa}}\setminus\bigcup_{\widehat{\mathbf p}\in\widehat{\mathcal P}^{(k-1)}}\mathcal B(\widehat{\mathbf p},d_{\rm{NMS}})$.

{For global search, let $\mathbf e_1$ be the first column of $\mathbf I_{N_{\rm{rx}}^{(\rm{A})}}$ and define $\mathbf S_{\mathrm{A},1}=\mathbf e_1^T\otimes\mathbf I_G$. It extracts $\tilde{\boldsymbol\phi}_{\mathrm{A},1}(\mathbf p)=\mathbf S_{\mathrm{A},1}\tilde{\boldsymbol\phi}_{\mathrm{A,ST}}(\mathbf p)$ and $\mathbf r_{\mathrm{A},1}^{(k-1)}=\mathbf S_{\mathrm{A},1}\mathbf r_{\mathrm{A}}^{(k-1)}$. Their normalized versions follow the definition above. The coarse spectrum is represented as}
\begin{equation}
\label{jk}
\begin{aligned}
J^{(k)}(\mathbf{p}_{\nu})
=
\left|
\bar{\boldsymbol{\phi}}_{\mathrm{A,1}}^{H}(\mathbf{p}_{\nu})
\bar{\mathbf{r}}_{\mathrm{A},1}^{(k-1)}
\right|
+
\sum_{i=1}^{I}
\left|
\bar{\boldsymbol{\phi}}_{i}^{H}(\mathbf{p}_{\nu})
\bar{\mathbf{r}}_{i}^{(k-1)}
\right|.
\end{aligned}
\end{equation}
For a threshold $\gamma_{\rm{th}}>0$, if $\max_{\mathbf{p}_{\nu}\in\mathcal{P}_{\rm{coa}}^{(k)}}
J^{(k)}(\mathbf{p}_{\nu})<\gamma_{\rm{th}}$,
the current global acquisition returns no candidate, and Algorithm~1 applies the corresponding continuation or termination rule.
Otherwise, NMS sequentially retains up to the prescribed maximum of $M_{\rm{c}}$ spatially separated peaks, where $a\in\{1,\dots,M_{\rm{c}}\}$ indexes the extraction order, which is shown as
\begin{equation}
\label{eq:nms_candidates}
\mathbf{p}_{\rm{c},a}^{(k)}
=
\underset{
\mathbf{p}_{\nu}\in
\mathcal{P}_{\rm{coa}}^{(k)}
\setminus
\displaystyle\bigcup_{a'=1}^{a-1}
\mathcal{B}
\left(
\mathbf{p}_{\rm{c},a'}^{(k)},
d_{\rm{NMS}}
\right)}
{\arg\max}\;
J^{(k)}(\mathbf{p}_{\nu}),
\end{equation}
where $a'$ indexes previously retained peaks. Extraction stops when the remaining set is empty or its largest score is below $\gamma_{\rm{th}}$, yielding the actual candidate count $M_{\rm{c}}^{(k)}\leq M_{\rm{c}}$. These candidates form $\mathcal C^{(k)}=\{\mathbf p_{\rm{c},a}^{(k)}\}_{a=1}^{M_{\rm{c}}^{(k)}}$.

\subsubsection{Candidate-Wise Fine Search and Off-Grid Refinement}
Let $d_{\rm{sea}}>0$ be the initial search radius and $d_{\rm{sea}}^{(\ell)}$ the radius at refinement level $\ell\in\{0,\dots,L_{\rm{exp}}\}$, with $d_{\rm{sea}}^{(0)}=d_{\rm{sea}}$ and $L_{\rm{exp}}$ denoting the maximum expansion level. Because prediction or coarse-grid errors may place the spectral peak outside the initial ROI, $\ell$ indexes successive ROI expansions triggered by a boundary-adjacent local maximum.
For each $\mathbf p_{\rm{c},a}^{(k)}\in\mathcal C^{(k)}$, the ROI is represented as
${\mathcal{B}_{a}^{(k,\ell)}
=\mathcal B\left(\mathbf p_{\rm{c},a}^{(k)},d_{\rm{sea}}^{(\ell)}\right)}$.
Then, the fine-search spectrum explicitly uses the full BS-A ST atom as
\begin{equation}
\label{eq:st_fine_spectrum}
J_{\rm{ST}}^{(k)}(\mathbf p)=\left|\bar{\boldsymbol\phi}_{\mathrm{A,ST}}^H(\mathbf p)\bar{\mathbf r}_{\mathrm{A}}^{(k-1)}\right|+\sum_{i=1}^{I}\left|\bar{\boldsymbol\phi}_{i}^H(\mathbf p)\bar{\mathbf r}_{i}^{(k-1)}\right|.
\end{equation}
At level $\ell$, let $\mathcal P_{\rm{loc},a}^{(k,\ell)}\subset\mathcal B_a^{(k,\ell)}$ denote the local 3D grid. Its maximizer is shown as
\begin{equation}
\label{plocx}
\mathbf{p}_{\rm{loc},a}^{\star,(\ell)}
=
\underset{
\mathbf{p}\in\mathcal{P}_{\rm{loc},a}^{(k,\ell)}
}
{\arg\max}
\;
J_{\rm{ST}}^{(k)}(\mathbf{p}).
\end{equation}
Define the normalized boundary metric as
\begin{equation}
\zeta_a^{(k,\ell)}=\frac{\|\mathbf p_{\rm{loc},a}^{\star,(\ell)}-\mathbf p_{\rm{c},a}^{(k)}\|_2}{d_{\rm{sea}}^{(\ell)}}.
\end{equation}
If $\zeta_a^{(k,\ell)}\geq\rho_{\rm{b}}$, the radius is updated as $d_{\rm{sea}}^{(\ell+1)}=\kappa_{\rm{exp}}d_{\rm{sea}}^{(\ell)}$ until the peak is no longer boundary-adjacent or $\ell=L_{\rm{exp}}$. Here, $\rho_{\rm{b}}\in(0,1]$ is the boundary threshold and $\kappa_{\rm{exp}}>1$ is the radius-expansion factor. The grid position obtained when expansion terminates is denoted $\mathbf p_{\rm{loc},a}^{\star}$.

The off-grid refinement is applied to reduce the residual spatial discretization error.
For axis-wise parabolic interpolation, let $\varsigma$ denote the axis displacement, $\mathbf e_x=[1,0,0]^T$ the X-axis unit vector, and $\epsilon>0$ the interpolation step. Define $f_{x,a}^{(k)}(\varsigma)=J_{\rm{ST}}^{(k)}(\mathbf p_{\rm{loc},a}^{\star}+\varsigma\mathbf e_x)$. The X-axis offset is
\begin{equation}
	\label{deltax}
	\Delta x_a
	=
	\frac{\epsilon}{2}
	\frac{
	f_{x,a}^{(k)}(-\epsilon)
	-
	f_{x,a}^{(k)}(\epsilon)
	}{
	f_{x,a}^{(k)}(\epsilon)
	+
	f_{x,a}^{(k)}(-\epsilon)
	-
	2f_{x,a}^{(k)}(0)
	}.
\end{equation}
The offsets $\Delta y_a$ and $\Delta z_a$ follow analogously, yielding
\begin{equation}
	\label{pfine}
	\widehat{\mathbf{p}}_{\rm{fine},a}^{(k)}
	=
	\mathbf{p}_{\rm{loc},a}^{\star}
	+
	[\Delta x_a,\Delta y_a,\Delta z_a]^T.
\end{equation}

\vspace{-2mm}
\subsubsection{Candidate Verification and Interference Cancellation}

After the candidate-wise ST refinement, each position must be verified before being formally introduced into the SIC. For the $a$-th candidate, its projected atom at BS $i'$ is defined as
\begin{equation}
\label{eq:candidate_atom}
\mathbf{q}_{i',a}^{(k)}
=
\begin{cases}
\tilde{\boldsymbol{\phi}}_{\mathrm{A,ST}}
\left(
\widehat{\mathbf{p}}_{\rm{fine},a}^{(k)}
\right), & i'=\mathrm{A},\\[1mm]
\tilde{\boldsymbol{\phi}}_{i'}
\left(
\widehat{\mathbf{p}}_{\rm{fine},a}^{(k)}
\right), & i'\in\{1,\ldots,I\}.
\end{cases}
\end{equation}
The candidate atom is temporarily appended to the current active dictionary as
\begin{equation}
\label{eq:temporary_dictionary}
\widetilde{\boldsymbol{\Phi}}_{i',a}^{(k)}
=
\left[
\widetilde{\boldsymbol{\Phi}}_{i'}^{(k-1)},
\mathbf{q}_{i',a}^{(k)}
\right].
\end{equation}
Using this temporary dictionary, the corresponding LS coefficients are defined as
\begin{equation}
\label{eq:temporary_ls}
\widetilde{\boldsymbol{\alpha}}_{i',a}^{(k)}
=
\left[
\left(\widetilde{\boldsymbol{\Phi}}_{i',a}^{(k)}\right)^H
\widetilde{\boldsymbol{\Phi}}_{i',a}^{(k)}
\right]^{-1}
\left(\widetilde{\boldsymbol{\Phi}}_{i',a}^{(k)}\right)^H
\tilde{\mathbf u}_{i'}^{\rm{res}}.
\end{equation}
The corresponding transient residual is shown as
\begin{equation}
\label{eq:temporary_residual}
\widetilde{\mathbf r}_{i',a}^{(k)}
=
\tilde{\mathbf u}_{i'}^{\rm{res}}
-
\widetilde{\boldsymbol{\Phi}}_{i',a}^{(k)}
\widetilde{\boldsymbol{\alpha}}_{i',a}^{(k)}.
\end{equation}
To quantify how well each candidate explains the multi-BS observations, the aggregate residual energies before and after its temporary inclusion are compared.
The joint residual-reduction ratio is defined as
\begin{equation}
\label{eq:candidate_verification}
\xi_a^{(k)}
=
\frac{
\displaystyle
\sum_{i'\in\mathcal I}
\left\|
\mathbf{r}_{i'}^{(k-1)}
\right\|_2^2
-
\displaystyle
\sum_{i'\in\mathcal I}
\left\|
\widetilde{\mathbf{r}}_{i',a}^{(k)}
\right\|_2^2
}{
\displaystyle
\sum_{i'\in\mathcal I}
\left\|
\mathbf{r}_{i'}^{(k-1)}
\right\|_2^2
}.
\end{equation}
For the residual-reduction threshold $\xi_{\rm{th}}\in(0,1)$, only candidates satisfying
$\xi_a^{(k)}\geq\xi_{\rm{th}}$
are retained in the valid candidate set $\mathcal{V}^{(k)}
=\left\{a:\xi_a^{(k)}\geq\xi_{\rm{th}}\right\}$.
If $\mathcal V^{(k)}=\emptyset$, the corresponding fallback or termination procedure is applied as summarized in Algorithm~1.
For a nonempty $\mathcal{V}^{(k)}$, the final candidate index is shown as
\begin{equation}
\label{eq:accepted_index}
a^{\star}
=
\underset{a\in\mathcal{V}^{(k)}}{\arg\max}\;
\xi_a^{(k)},
\end{equation}
and the corresponding accepted position is finally obtained as
\begin{equation}
\label{eq:accepted_position}
\widehat{\mathbf{p}}_{\rm{fine}}^{(k)}
=
\widehat{\mathbf{p}}_{\rm{fine},a^{\star}}^{(k)}.
\end{equation}
Once accepted, the candidate is used for both track management and SIC. If the accepted position remains associated with the scheduled track $\eta_k$, it updates that track. For the associated track, the superscript $+$ denotes its updated state, the position is replaced by $\mathbf p_{\rm{last},\eta_k}^{+}=\widehat{\mathbf p}_{\rm{fine}}^{(k)}$, and the velocity is updated using the smoothing factor $\rho_{\rm{s}}\in[0,1]$ as
\begin{equation}
\label{vv}
\mathbf{v}_{\rm{est},\eta_k}^{+}
=
\rho_{\rm{s}}\mathbf{v}_{\rm{est},\eta_k}
+
\left(1-\rho_{\rm{s}}\right)
\frac{
\widehat{\mathbf{p}}_{\rm{fine}}^{(k)}
-
\mathbf{p}_{\rm{last},\eta_k}
}{
\Delta T
}.
\end{equation}
Otherwise, it initializes a new track with the accepted position and zero velocity.

For SIC, the temporary dictionary, LS coefficients, and residual associated with $a^{\star}$ in \eqref{eq:temporary_dictionary}--\eqref{eq:temporary_residual} are retained as the active quantities $\widetilde{\boldsymbol{\Phi}}_{i'}^{(k)}$, $\widehat{\boldsymbol{\alpha}}_{i'}^{(k)}$, and $\mathbf r_{i'}^{(k)}$, respectively.
Therefore, only candidates that consistently reduce the multi-BS residual energy are allowed to enter the SIC process.
The complete procedure is summarized in Algorithm~1.

\color{black}
\begin{algorithm}[t]
	\caption{Coarse-to-Fine Multi-Static Cooperative UAV Localization}
	\label{alg:localization_sic}
	\footnotesize
	\begin{algorithmic}[1]
		
		\Require $\{\tilde{\mathbf u}_{i'}^{\rm{obs}},\boldsymbol\phi_{i'}^{\rm{DP}}\}_{i'\in\mathcal I}$,
		$\mathcal P_{\rm{coa}}$, the parameters defined above, and established track states
		$\{\mathbf p_{\rm{last},q},\mathbf v_{\rm{est},q}\}_{q=1}^{Q_{\rm{trk}}}$.
		\Ensure Accepted target-position set $\widehat{\mathcal P}$ and updated track states.

		\State \textbf{Phase 1: DPI Suppression and Initialization}
		\State Compute $\tilde{\mathbf u}_{i'}^{\rm{res}}$ via \eqref{u_res_local} and set
		$\mathbf r_{i'}^{(0)}\leftarrow\tilde{\mathbf u}_{i'}^{\rm{res}}$,
		$\widetilde{\boldsymbol\Phi}_{i'}^{(0)}\leftarrow\emptyset$, and
		$\widehat{\mathcal P}^{(0)}\leftarrow\emptyset$ for all $i'\in\mathcal I$.

		\State \textbf{Phase 2: Kinematic Prediction and Priority Ordering}
		\State Predict and score all established tracks via \eqref{ppred} and \eqref{eq},
		then sort them to obtain $\{\eta_k\}_{k=1}^{K_{\max}}$, with $\eta_k=0$
		after all tracks have been scheduled.

		\State \textbf{Phase 3: Robust Target Extraction and SIC}
		\For{$k\in\{1,\ldots,K_{\max}\}$}
			\If{$\eta_k\neq 0$}
				\State Set $\mathcal C^{(k)}\leftarrow\{\mathbf p_{\rm{pred},\eta_k}\}$.
			\Else
				\State Form $\mathcal P_{\rm{coa}}^{(k)}$ and obtain $\mathcal C^{(k)}$ by
				thresholding \eqref{jk} and applying \eqref{eq:nms_candidates}.
			\EndIf

			\State Refine all centers in $\mathcal C^{(k)}$, including ROI expansion, via \eqref{eq:st_fine_spectrum}--\eqref{pfine}. Verify the refined candidates
			via \eqref{eq:candidate_atom}--\eqref{eq:candidate_verification} and form
			$\mathcal V^{(k)}$.

			\If{$\mathcal V^{(k)}=\emptyset\ \text{and}\ \eta_k\neq 0$}
				\State Form $\mathcal P_{\rm{coa}}^{(k)}$, perform one global acquisition,
				and repeat the refinement and verification via
				\eqref{eq:st_fine_spectrum}--\eqref{eq:candidate_verification}.
			\EndIf

			\If{$\mathcal V^{(k)}=\emptyset$}
				\State Set $\mathbf r_{i'}^{(k)}\leftarrow\mathbf r_{i'}^{(k-1)}$,
				$\widetilde{\boldsymbol\Phi}_{i'}^{(k)}\leftarrow
				\widetilde{\boldsymbol\Phi}_{i'}^{(k-1)}$, and
				$\widehat{\mathcal P}^{(k)}\leftarrow\widehat{\mathcal P}^{(k-1)}$.
				If $\eta_k=0$, \textbf{break}. Otherwise, \textbf{continue}.
			\EndIf

			\State Select $a^{\star}$ and $\widehat{\mathbf p}_{\rm{fine}}^{(k)}$ via
			\eqref{eq:accepted_index} and \eqref{eq:accepted_position}.
			\State Update the scheduled track via \eqref{vv} only when the accepted
					position remains associated with it. Otherwise, initialize a new track.
			\State Retain the temporary dictionary, LS coefficients, and residual for
			$a^{\star}$ in \eqref{eq:temporary_dictionary}--\eqref{eq:temporary_residual}
			as $\widetilde{\boldsymbol\Phi}_{i'}^{(k)}$,
			$\widehat{\boldsymbol\alpha}_{i'}^{(k)}$, and $\mathbf r_{i'}^{(k)}$, and set
			$\widehat{\mathcal P}^{(k)}\leftarrow\widehat{\mathcal P}^{(k-1)}
			\cup\{\widehat{\mathbf p}_{\rm{fine}}^{(k)}\}$.
		\EndFor

		\State \Return the latest accepted-position set as $\widehat{\mathcal P}$ and
		updated track states.
		\color{black}
	\end{algorithmic}
\end{algorithm}

\vspace{-5mm}
\subsection{Computational Complexity Analysis}
	{
To evaluate efficiency, we focus on pure localization-search operations.
Let $N_{\rm coa}=G(I+1)$ and $N_{\rm fine}=G(N_{\rm rx}^{(\rm A)}+I)$ denote the observation dimensions used by coarse acquisition and full ST refinement, while $N_{\rm c}$ and $N_{\rm loc}$ denote the numbers of global coarse and candidate-wise local grids, respectively, with $N_{\rm loc}$ including possible ROI expansion. Evaluating one local grid requires $\mathcal O(N_{\rm fine})$ operations, so refining one candidate costs $\mathcal O(N_{\rm fine}N_{\rm loc})$. In global acquisition, the coarse search uses the first BS-A receive antenna and the $I$ distributed BSs, requiring $\mathcal O(N_{\rm coa}N_{\rm c})$ operations, followed by at most $M_{\rm c}$ local refinements. If prediction fails and global fallback is triggered, at most $M_{\rm c}+1$ candidates are refined. Therefore, over at most $K_{\max}$ attempts, a conservative upper bound is
$\mathcal O\!\Big(K_{\max}\big[N_{\rm coa}N_{\rm c}+N_{\rm fine}(M_{\rm c}+1)N_{\rm loc}\big]\Big)$. However, during stable tracking, a prediction-based candidate that passes verification bypasses global acquisition. Updating $Q$ verified tracks in one frame therefore has dominant complexity $\mathcal O(QN_{\rm fine}N_{\rm loc})$.}

\vspace{-3mm}
\section{Simulation Results}
To evaluate the proposed MIMO UW-OCDM ISAC system, simulations are conducted for both communication and sensing.
As depicted in Fig.~\ref{Fig_sensing}, BS-A transmits UW-OCDM to a UAV, which may also be one of the sensing targets, while BS-A and $I=3$ geographically distributed sensing BSs, denoted by BS-1, BS-2, and BS-3, receive echoes from the UAV targets for cooperative multi-static sensing.
	The system operates at a carrier frequency $f_{\rm{c}} = 5$ GHz with a bandwidth $B = 40$ MHz (i.e., a baseband sampling rate $f_{\rm{s}} = 40$ MHz), employing $16$-QAM for payload data. Performance is evaluated over different signal-to-noise ratios (SNRs).
	The UW-OCDM symbol length is $N = 2048$, and the QPSK UW of length $N_{\rm{UW}} = 256$ is realized using $N_{\rm{g}} = 384$ equispaced sub-chirps for each transmit antenna.
With $L_{\max}=64$, the receiver retains an ISI-free observation window of length $G=N_{\rm{UW}}-L_{\max}=192$. The delay margin provides a path-length budget of $R_{\max}=cL_{\max}/f_{\rm{s}}=480$ m.

\vspace{-3mm}
\subsection{UAV Communication Performance Evaluation}
For the communication link, the system employs a Rician fading channel with $K_{\rm{Ri}} = 10$ dB, comprising one LoS path and $N_{\rm{path}} = 6$ NLoS scattering paths. 
The BS-A transmitter and UAV receiver are equipped with $N_{\rm{tx}}^{(\rm{A})} = 2$ and $N_{\rm{rx}}^{(\rm{U})} = 4$ antennas, respectively. 
To simulate a highly dynamic GA scenario, the velocity governing the LoS Doppler shift $f_{\rm{D,LoS}}$ is fixed at $100$ km/h, while the NLoS equivalent velocities are distributed between $0$ and $100$ km/h. 
The oscillator-induced carrier frequency offset is assumed perfectly compensated to focus on mobility-induced Doppler.
Specifically, two adjacent UWs, i.e., $X_{\rm{num}} = 2$ and $D = 1$, are used for Doppler estimation and MMV-OMP CE.
Unless otherwise specified, all communication simulations adopt parameters as shown above.

\subsubsection{UW Sub-Chirp Power Constraints}
To evaluate the effect of limiting large UW-generating sub-chirp coefficients, we compare direct nonlinear clipping (Clip), overall linear scaling (Scale), and projected gradient descent (PGD) \cite{Bertsekas1976PGD}.
{These are analyzed across truncation thresholds $A_{\rm{th}} \in \{2, 4, 8, 16\}$, defined as multiples of the normalized constellation's peak amplitude.}
{The Clip strategy nonlinearly restricts any threshold-exceeding UW-generating sub-chirp amplitude to $A_{\rm{th}}$.}
The total energy of this clipped sequence establishes a baseline data-to-UW power ratio. 
To guarantee a fair comparison, the remaining two strategies strictly adhere to this baseline. 
Specifically, the Scale strategy uniformly attenuates the overall amplitude of the unconstrained sub-chirps to match this exact energy, while the PGD strategy constrains its $\ell_2$-norm not to exceed that of the clipped sequence.
{Consequently, Scale matches the clipped-sequence energy, while PGD does not exceed it for any $A_{\rm{th}}$.}

\begin{figure}[!t]
	\captionsetup{font={footnotesize}, name = {Fig.}, labelsep = period}
	\centering
	\includegraphics[width=8cm]{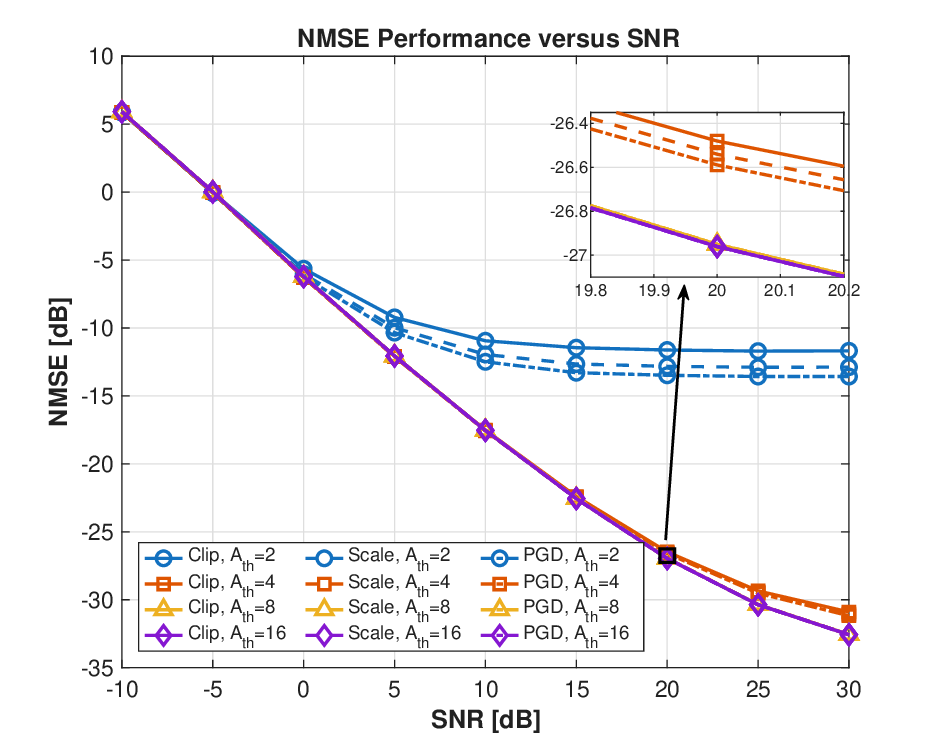}
	\caption{CE normalized mean-square error (NMSE) performance of different power-constraint strategies.}
	\label{Fig_NMSE_Power}
	\vspace*{-5mm}
\end{figure}
\begin{figure}[!t]
	\captionsetup{font={footnotesize}, name = {Fig.}, labelsep = period}
	\centering
	\includegraphics[width=8cm]{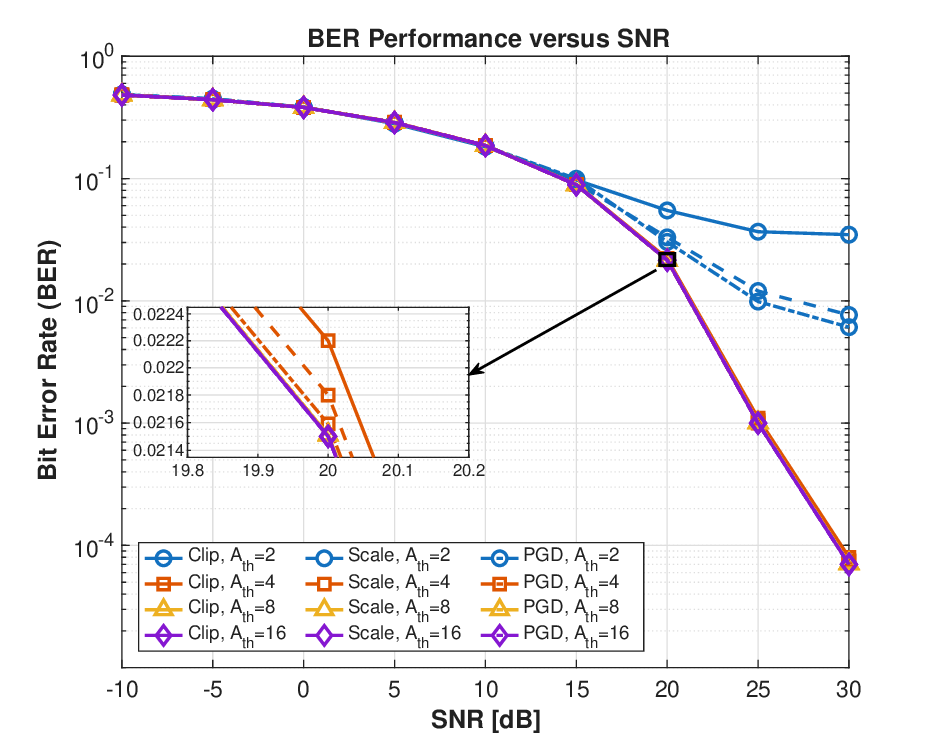}
	\caption{Data demodulation bit error rate (BER) performance of different power-constraint strategies.}
	\label{Fig_BER_Power}
	\vspace*{-7mm}
\end{figure}

{As Fig.~\ref{Fig_NMSE_Power} illustrates, the severe amplitude limit at $A_{\rm{th}}=2$ degrades CE performance, yielding a higher NMSE.}
The Clip strategy suffers the most from severe distortion, while the Scale and PGD strategies offer slight mitigation. 
{Correspondingly, Fig.~\ref{Fig_BER_Power} shows that the BER for all strategies improves and converges as $A_{\rm{th}}$ increases.}
{At $A_{\rm{th}}=16$, the amplitude constraint is effectively inactive under the evaluated settings, yielding the best performance.}
Within the evaluated settings, relaxing the amplitude constraint provides the best NMSE and BER performance. Hence, the solution in \eqref{psedo} is adopted without additional clipping.
\begin{figure}[!t]
	\captionsetup{font={footnotesize}, name = {Fig.}, labelsep = period}
	\centering
	\includegraphics[width=8cm]{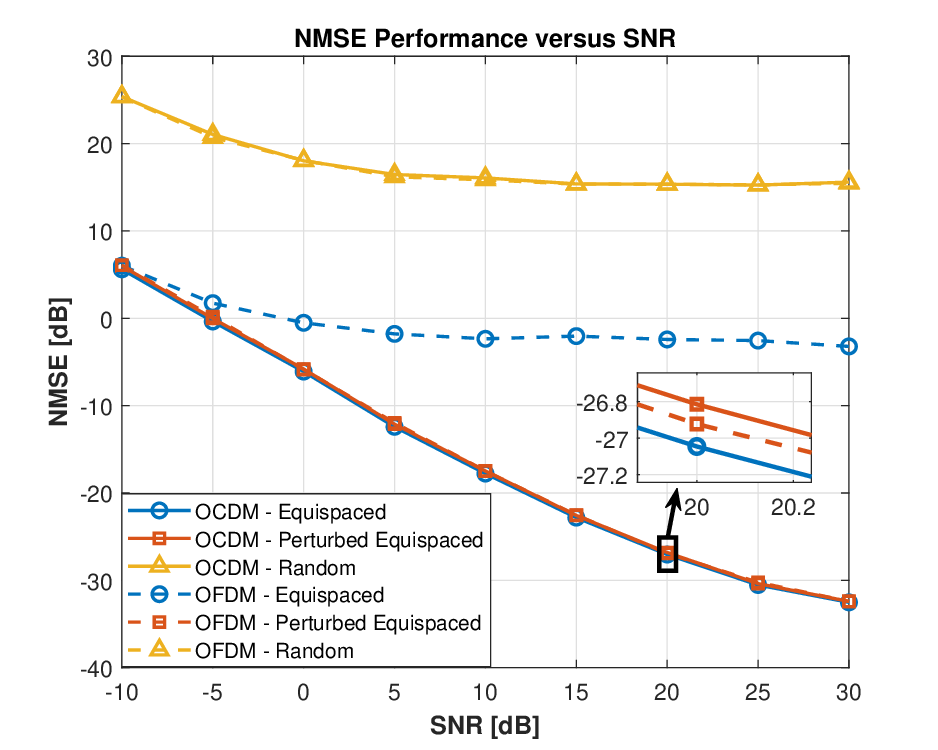}
	\caption{CE NMSE performance of different subcarrier/sub-chirp index allocation strategies.}
	\label{Fig_NMSEjitter}
	\vspace*{-5mm}
\end{figure}
\begin{figure}[!t]
	\captionsetup{font={footnotesize}, name = {Fig.}, labelsep = period}
	\centering
	\includegraphics[width=8cm]{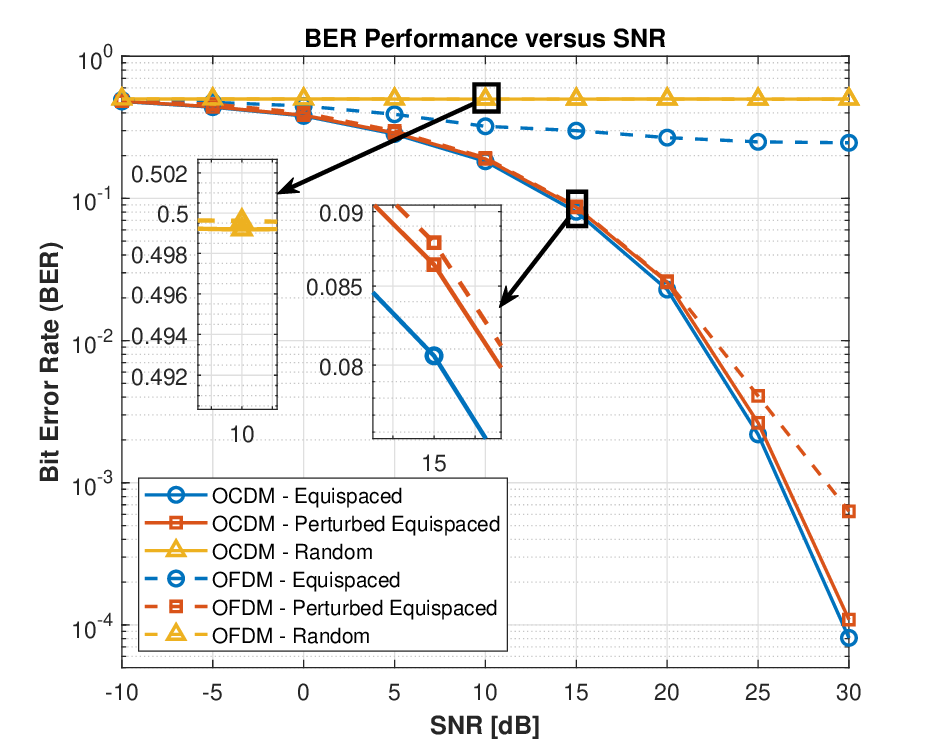}
	\caption{Data demodulation BER performance of different subcarrier/sub-chirp index allocation strategies.}
	\label{Fig_BERjitter}
	\vspace*{-3mm}
\end{figure}	
\subsubsection{UW Resource Index Allocation}
We investigate equispaced, pseudo-random (perturbed-equispaced), and fully random UW-generating resource allocations. UW-OCDM and UW-OFDM use the same UW, resource count, and average transmit power.
As shown in Figs.~\ref{Fig_NMSEjitter} and \ref{Fig_BERjitter}, fully random allocation severely degrades the CE NMSE and BER of both UW-OCDM and UW-OFDM because its clusters and large gaps worsen the conditioning of the UW-generation matrix. The resulting larger mutually cancelling coefficients increase redundant-power consumption and reduce the payload power under equal total transmit power.
The ideal UW autocorrelation is identical across allocations because the UW is fixed, so it cannot explain these performance differences. PAPR instead characterizes the complete transmit symbol and may vary with the generating coefficients, but it is not the criterion used to select the allocation.
The structured allocations affect the waveforms differently. In UW-OFDM, equispaced indices \cite{UW-OFDM} generate periodic time-domain repetitions and correlation ambiguity through the inverse DFT. Pseudo-random perturbations break this periodicity while retaining near-uniform coverage. 
In UW-OCDM, the index-dependent quadratic phase rotation introduced by the DFnT suppresses this repetitive behavior even with equispaced sub-chirps.
Consequently, equispaced allocation is adopted for UW-OCDM owing to its favorable performance, conditioning, and power efficiency. Additionally, all subsequent UW-OFDM simulations use pseudo-random allocation.

\begin{figure}[!t]
	\captionsetup{font={footnotesize}, name = {Fig.}, labelsep = period}
	\centering
	\includegraphics[width=8cm]{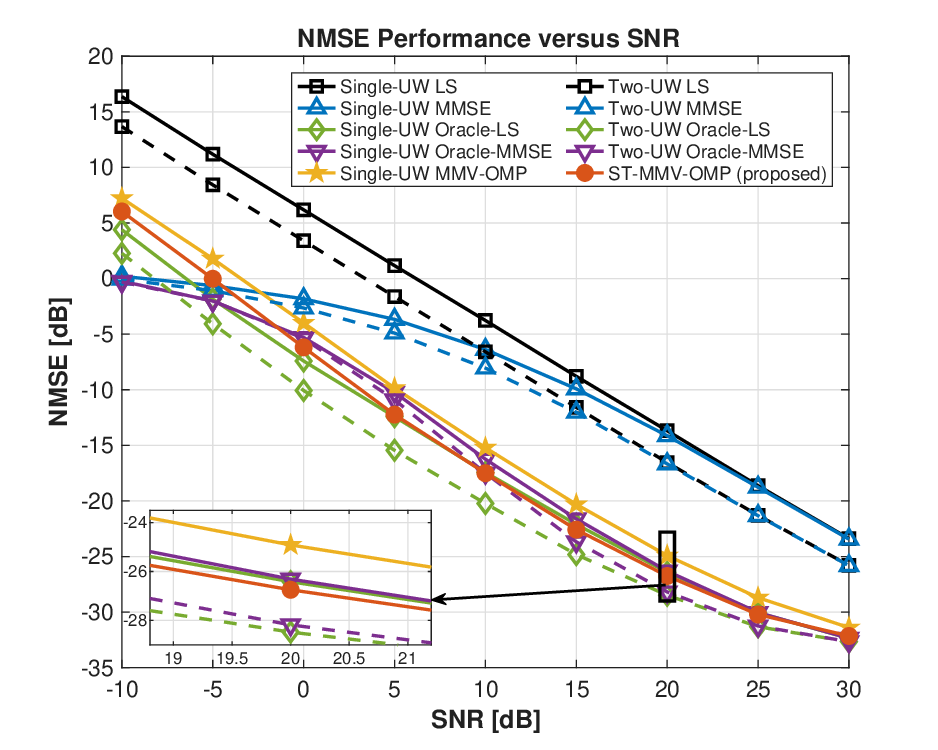}
	\caption{CE NMSE performance of different CE schemes.}
	\label{Fig_NMSE5}
	\vspace*{-5mm}
\end{figure}
\begin{figure}[!t]
	\captionsetup{font={footnotesize}, name = {Fig.}, labelsep = period}
	\centering
	\includegraphics[width=8cm]{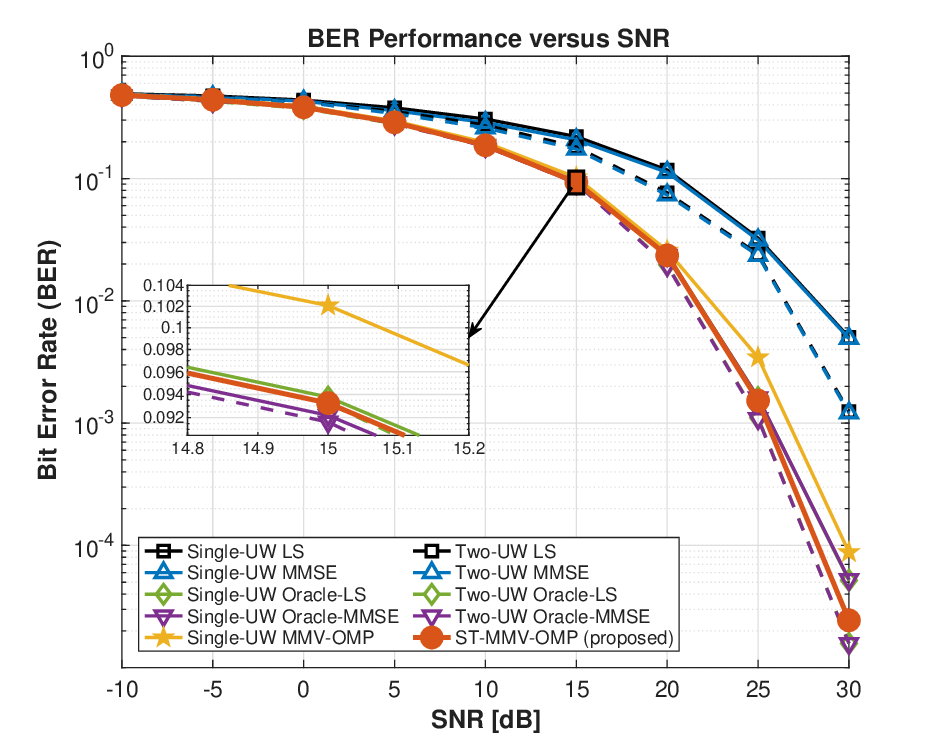}
	\caption{Data demodulation BER performance of different CE schemes.}
	\label{Fig_BER5}
	\vspace*{-5mm}
\end{figure}

\subsubsection{Channel Estimation Performance}
We compare the proposed two-UW spatio-temporal MMV-OMP (ST-MMV-OMP) scheme with a single-UW spatial MMV-OMP ablation.
The latter retains joint processing across receive antennas but excludes temporal coupling, thereby isolating the gain from the additional UW observation. 
Under each observation setting, full-dimensional LS and MMSE estimators and their known-support Oracle-LS and Oracle-MMSE counterparts are included as references.
Fig.~\ref{Fig_NMSE5} shows that two-UW ST-MMV-OMP achieves lower NMSE than its single-UW counterpart, confirming the benefit of temporal coupling. Under either setting, MMV-OMP outperforms the full-dimensional estimators. 
Fig.~\ref{Fig_BER5} shows the same hierarchy. 
The two-UW scheme achieves lower BER than its single-UW counterpart, and MMV-OMP descends faster than the full-dimensional estimators.
Its advantage over the single-UW oracle references results from the additional temporal observation.
Overall, temporal coupling and joint support recovery improve CE and symbol recovery.
	
\subsubsection{Performance under Different Velocities}
		\begin{figure}[!t]
	\captionsetup{font={footnotesize}, name = {Fig.}, labelsep = period}
	\centering
	\includegraphics[width=8cm]{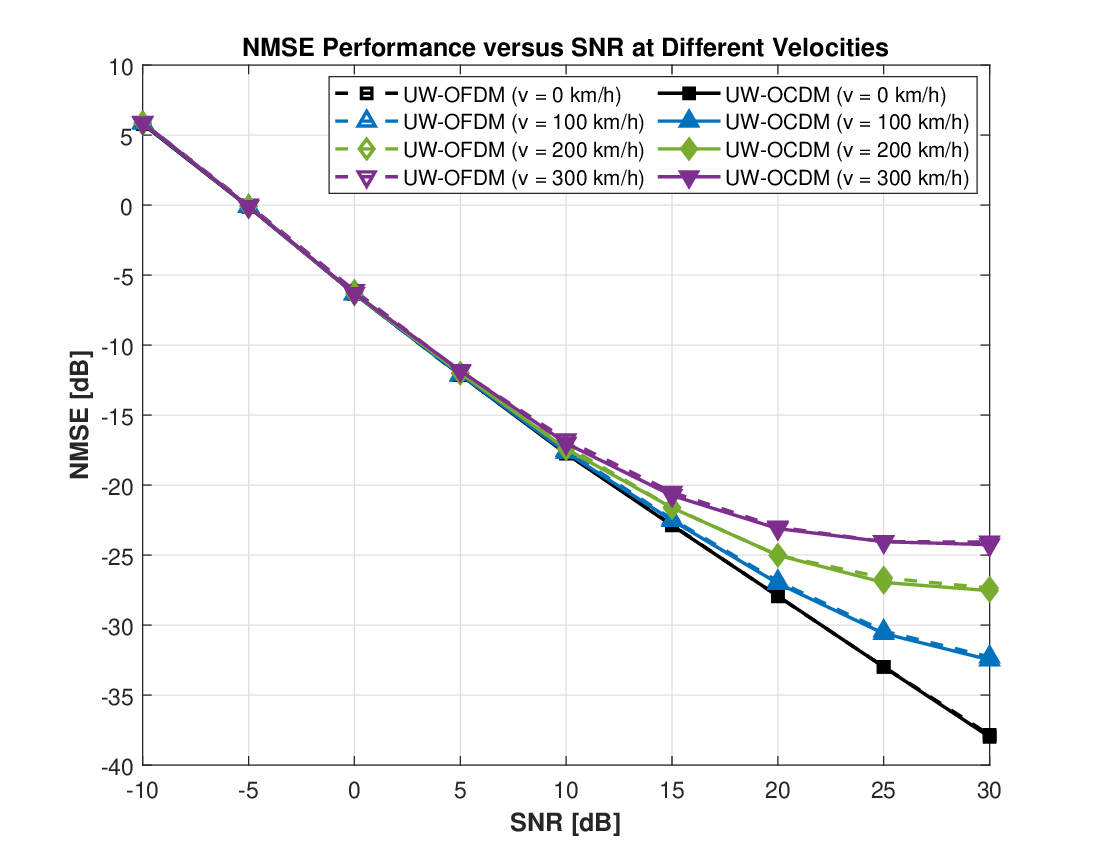}
	\caption{CE NMSE performance under different velocities.}
	\label{Fig_NMSEv}
	\vspace*{-5mm}
\end{figure}

\begin{figure}[!t]
	\captionsetup{font={footnotesize}, name = {Fig.}, labelsep = period}
	\centering
	\includegraphics[width=8cm]{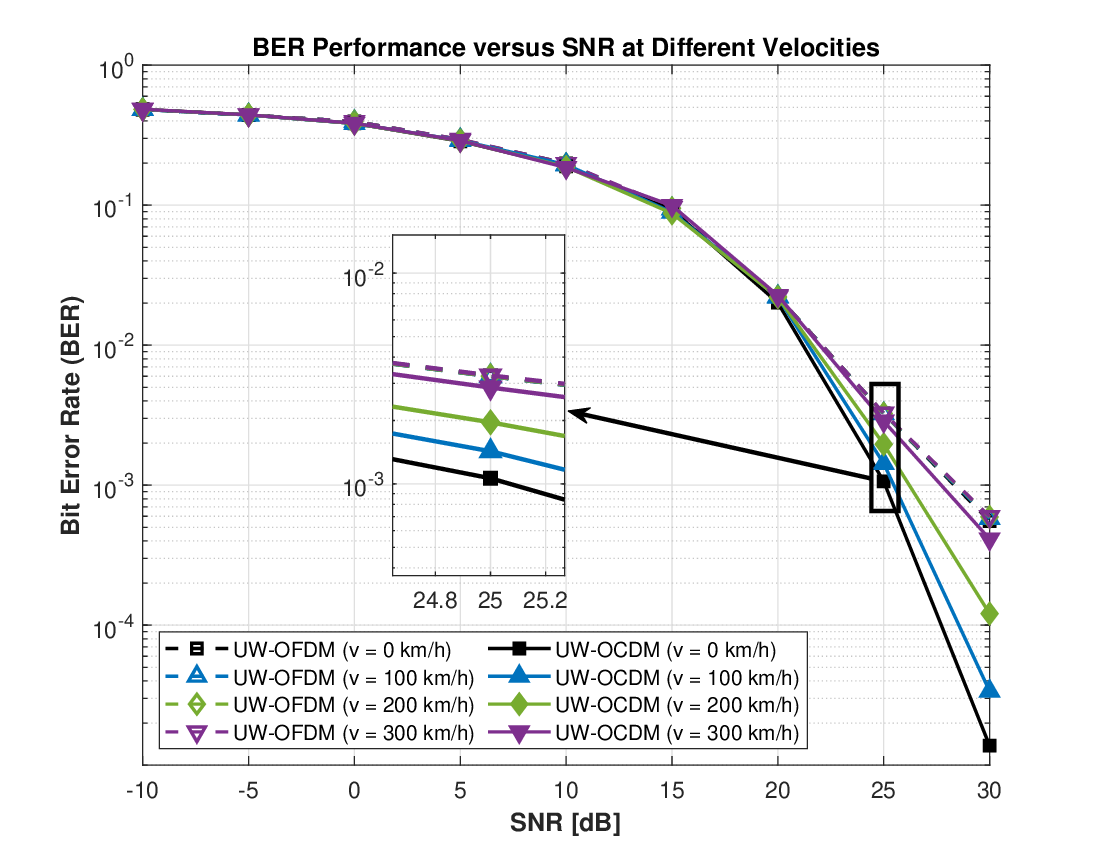}
	\caption{Data demodulation BER performance under different velocities.}
	\label{Fig_BERv}
	\vspace*{-7mm}
\end{figure}
To evaluate mobility robustness, we compare UW-OFDM and UW-OCDM at velocities from $0$ to $300$ km/h. As shown in Fig.~\ref{Fig_NMSEv}, at high SNRs, the NMSE gradually degrades with velocity because residual Doppler becomes more pronounced after dominant LoS compensation. 
The corresponding UW-OFDM and UW-OCDM curves remain close, indicating that both schemes use UW-assisted CE effectively.
Fig.~\ref{Fig_BERv} shows that both schemes maintain favorable BER across the evaluated velocities because the dominant LoS Doppler is estimated from two adjacent UWs and compensated before CE and data equalization. The tightly grouped UW-OFDM curves indicate small velocity-induced variation, although their high-SNR BER remains higher. UW-OCDM instead shows clearer high-SNR degradation with velocity because residual path-dependent Doppler remains after compensation. Despite this greater variation, UW-OCDM retains a lower high-SNR BER at corresponding velocities owing to the Doppler resilience of its chirp-based modulation. At $300$ km/h, its BER continues to decrease without a pronounced error floor.

\subsubsection{Waveform Performance Comparison}
	\begin{figure}[!t]
	\captionsetup{font={footnotesize}, name = {Fig.}, labelsep = period}
	\centering
	\includegraphics[width=8cm]{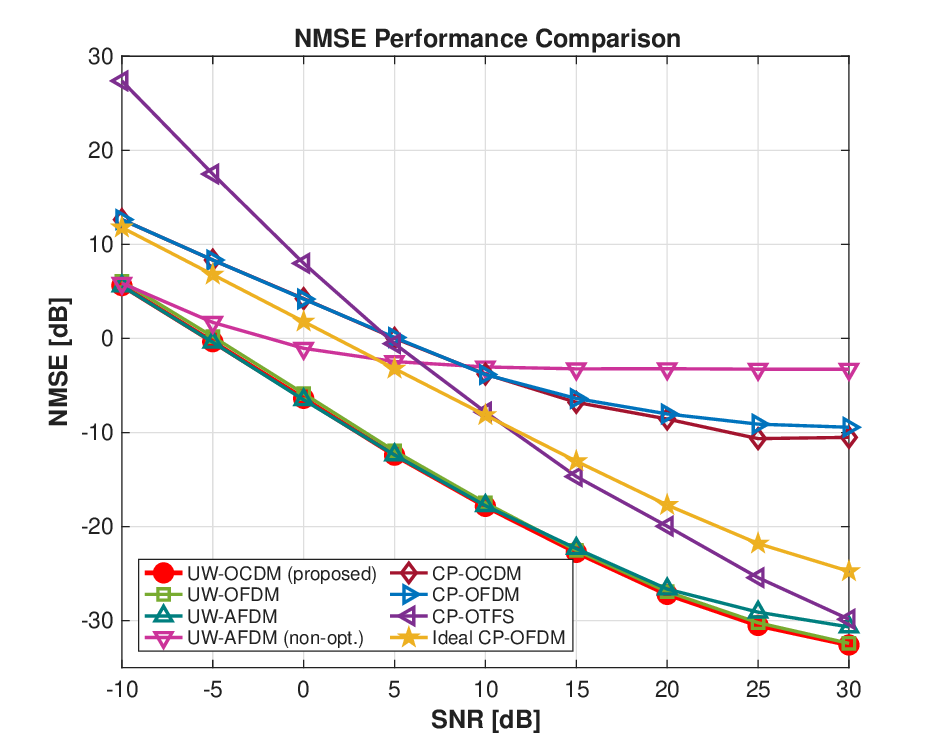}
	\caption{CE NMSE performance of the proposed UW-OCDM and representative waveform, ablation, and reference schemes.}
	\label{Fig_NMSEcomp}
	\vspace*{-5mm}
\end{figure}

\begin{figure}[!t]
	\captionsetup{font={footnotesize}, name = {Fig.}, labelsep = period}
	\centering
	\includegraphics[width=8cm]{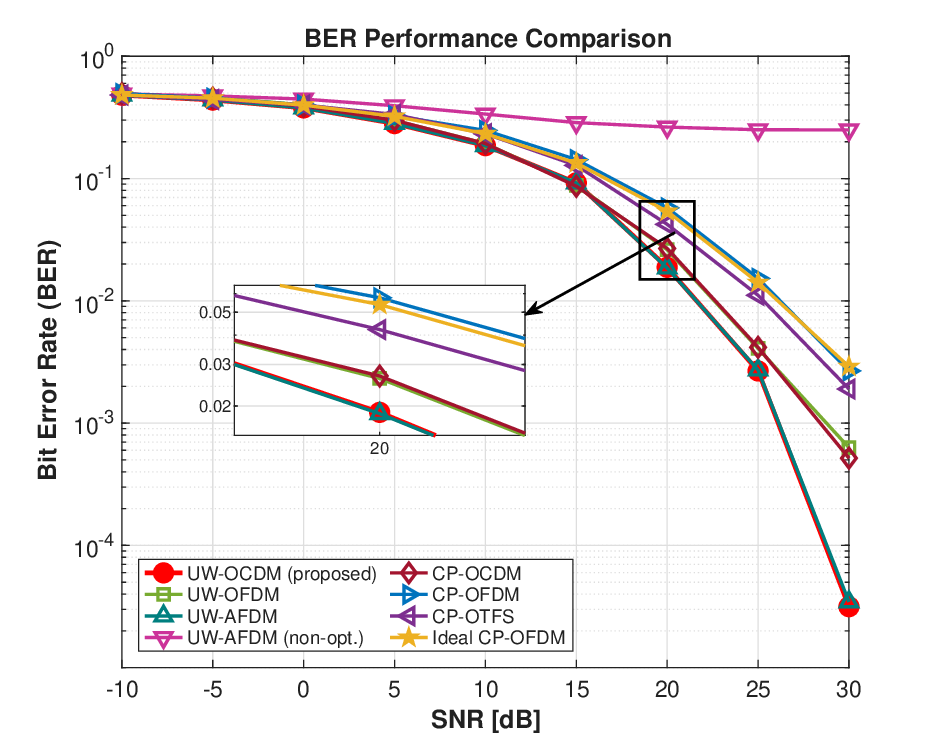}
	\caption{Data demodulation BER performance of the proposed UW-OCDM and representative waveform, ablation, and reference schemes.}
	\label{Fig_BERcomp}
	\vspace*{-7mm}
\end{figure}	
	{
		Figs.~\ref{Fig_NMSEcomp} and \ref{Fig_BERcomp} compare the CE NMSE and BER of the proposed UW-OCDM with representative waveform, ablation, and reference schemes.
		UW-AFDM \cite{Bemani2021AFDM} and its non-optimized variant are included to evaluate the effect of UW-based affine waveform design.
		CP-OCDM preserves OCDM modulation but replaces UW-based acquisition and CE with CP- and pilot-based processing, providing a direct ablation of the UW contribution.
	CP-OFDM and CP-OTFS serve as conventional and delay--Doppler baselines, respectively, while ideal CP-OFDM provides an upper reference under perfect synchronization.
	}

	{
Resource usage is matched as closely as each waveform structure permits. Each UW-based waveform reserves $N_{\rm{g}}=384$ of $N=2048$ transform-domain resources, leaving $N-N_{\rm{g}}=1664$ resources for payload and an effective data-resource ratio of $81.25\%$.
CP-OFDM, CP-OCDM, and ideal CP-OFDM use a $64$-sample CP and $332$ pilots to match this ratio.
CP-OTFS retains its native $256\times8$ grid, $64$-sample block CP, and $384$ pilot/guard elements, resulting in an effective rate only $3.03\%$ lower.
All schemes use the same bandwidth, sampling rate, 16-QAM payload modulation, MIMO configuration, channel realizations, and average transmit-sample power, while the UW-based schemes use the same QPSK UW. 
CP-OCDM and CP-OFDM employ delayed-CP correlation for synchronization, Doppler compensation, interpolated LS CE, and MMSE FDE. CP-OTFS adopts pilot correlation, delay--Doppler CE, and sparse LMMSE equalization.
	}
	{
		To relate AFDM and OCDM, we use $\mathbf{T}_{\rm{AFDM}}(c_1,c_2)=\boldsymbol{\Gamma}(c_1)\mathbf{F}_{N}^{H}\boldsymbol{\Gamma}(c_2)$, where $\boldsymbol{\Gamma}(c)=\operatorname{diag}\{e^{-\jmath2\pi c n^{2}}\}_{n=0}^{N-1}$. For even $N$, $\boldsymbol{\Lambda}^{H}=e^{\jmath\pi/4}\mathbf{T}_{\rm{AFDM}}(1/(2N),1/(2N))$, showing that OCDM is the symmetric AFDM point up to a common phase. 
		Therefore, we include AFDM with $c_1=1/(2N),c_2=\sqrt{2}/(2N)$ and a non-optimized setting $c_1=1/16,c_2=1/8$ as references.
	}

	{
		As shown in Figs.~\ref{Fig_NMSEcomp} and \ref{Fig_BERcomp}, among the UW-based waveforms, pseudo-random UW-OFDM attains CE NMSE close to UW-OCDM, but retains a higher BER. 
		UW-AFDM achieves nearly the same BER as UW-OCDM, but exhibits slightly higher NMSE in Fig.~\ref{Fig_NMSEcomp}.
		This can be explained by the fact that the dominant Doppler is compensated beforehand, while the adopted shared estimator and frequency-domain equalizer do not exploit DAFT-domain path separation.
		The non-optimized UW-AFDM produces clear error floors, indicating that AFDM parameters should be jointly selected with the UW constraint and receiver processing.
		CP-OCDM ablation then isolates the UW contribution because both schemes use the same OCDM modulation.
		UW-OCDM achieves substantially lower NMSE and lower BER, which is attributed to the deterministic UW reference and sparse MMV CE.
		Against the remaining references, UW-OCDM outperforms practical CP-OFDM and ideal CP-OFDM in both metrics.
		Although CP-OTFS improves steadily with SNR, UW-OCDM retains the best high-SNR NMSE and BER among the evaluated practical schemes.
		These results demonstrate that the OCDM and UW provide complementary gains.
	}

\begin{figure}[!t]
	\captionsetup{font={footnotesize}, name = {Fig.}, labelsep = period}
	\centering
	\includegraphics[width=8cm]{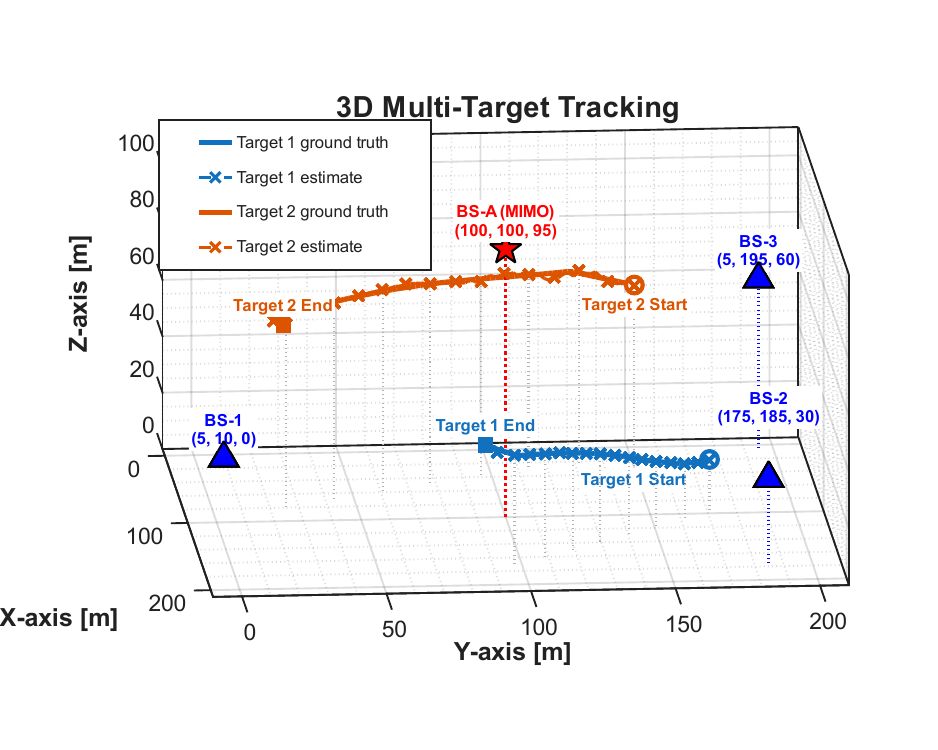}
	\caption{Ground-truth and estimated 3D UAV trajectories of the proposed localization method at an SNR of $30$ dB.}
	\label{Fig_3D}
	\vspace*{-7mm}
\end{figure}
	\subsection{UAV Localization Performance Evaluation}
	To evaluate sensing performance, we simulate a 3D target tracking scenario based on the aforementioned multi-static system. 
	BS-A employs transmit and receive ULAs with $d_{\rm A}=\lambda/2$ and respective element counts $N_{\rm{tx}}^{(\rm{A})}=2$ and $N_{\rm{rx}}^{(\rm{A})}=8$, whereas the $I=3$ distributed BSs are single-antenna receivers.
The BS coordinates\footnote{Multi-static localization accuracy depends on the BS-target geometry through the geometric dilution of precision and the conditioning of the localization problem \cite{Nguyen2016}. The adopted topology is representative rather than optimized. For another deployment, the sensing dictionary can be reconstructed using the updated BS coordinates, although the achieved accuracy will vary. Topology optimization is beyond the scope of this work.} (in meters) are $\mathbf p_{\mathrm A}=[100,100,95]^T$, $\mathbf p_1=[5,10,0]^T$, $\mathbf p_2=[175,185,30]^T$, and $\mathbf p_3=[5,195,60]^T$.
We generate $Q = 2$ dynamic UAV targets as the ground truth, each with an average radar cross-section (RCS) of $0.05~\mathrm{m}^2$ and random amplitude fluctuations.
Their initial coordinates are randomly distributed within a core detection zone spanning $0$ to $200$ m horizontally and $0$ to $100$ m vertically, with initial velocities ranging from $0$ to $30$ m/s.
	To generate nonuniform dynamic trajectories, collision avoidance and low-frequency sinusoidal wind drift are introduced. 
	The observation process spans $80$ tracking epochs with an update interval of $\Delta T = 0.1$ s.
	For the hierarchical target extraction, the coarse grid $\mathcal{P}_{\rm{coa}}$ employs step sizes of $\Delta x_{\rm{coa}} = \Delta y_{\rm{coa}} = 4.0$ m and $\Delta z_{\rm{coa}} = 3.0$ m, bounded by $200$ m along the X and Y axes, and $100$ m along the Z axis. 
	Furthermore, the kinematic velocity smoothing factor is set to $\rho_{\rm{s}} = 0.5$, and the axis-wise interpolation step is configured as $\epsilon = 0.2$ m.
	
\subsubsection{3D Trajectory Tracking Performance}
As shown in Fig.~\ref{Fig_3D}, the estimated positions closely follow both ground-truth trajectories over the $80$ tracking epochs without track loss. For visual clarity, markers on the estimated trajectories are displayed every $5$ tracking epochs, while all $80$ epochs are retained in the trajectory lines and performance evaluation. The close overlap is maintained through changes in direction and altitude, without evident error accumulation along either trajectory. The two estimated tracks also remain correctly associated from their marked start to end points despite simultaneous target motion and random RCS fluctuations. These results demonstrate that the proposed localization method can maintain continuous 3D multi-target tracking under the simulated motion and interference conditions.
	
\subsubsection{Ablation Study on Localization Performance}
\begin{figure}[!t]
	\captionsetup{font={footnotesize}, name = {Fig.}, labelsep = period}
	\centering
	\includegraphics[width=8cm]{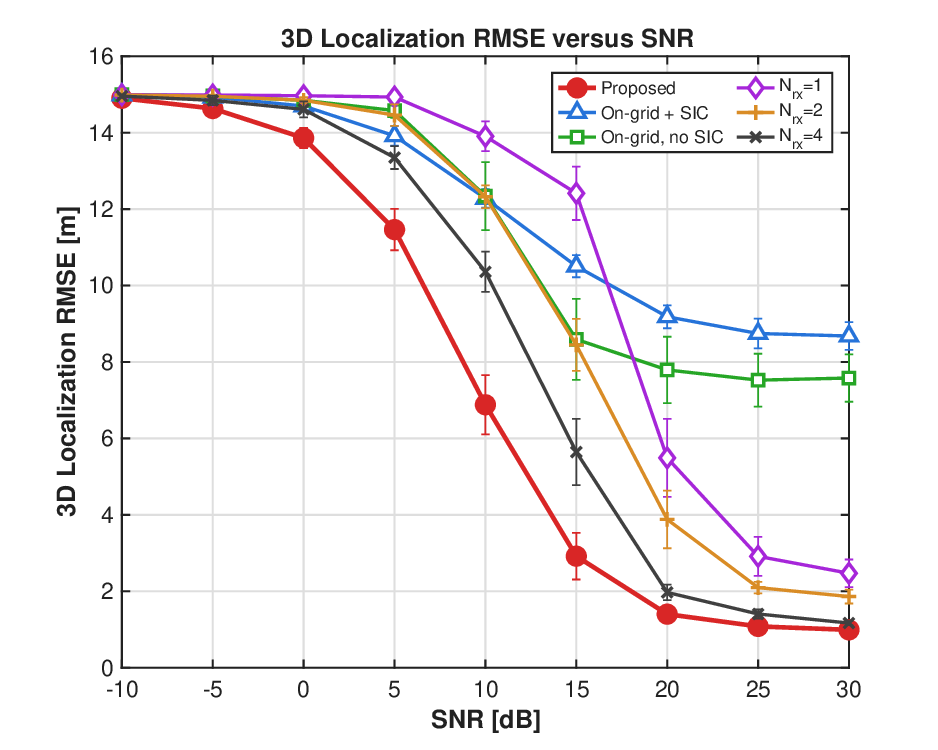}
	\caption{3D localization root-mean-square error (RMSE) of the proposed localization method and its off-grid, SIC, and receive-array ablations.}
	\label{Fig_RMSE_Ablation}
	\vspace*{-5mm}
\end{figure}

\begin{figure}[!t]
	\captionsetup{font={footnotesize}, name = {Fig.}, labelsep = period}
	\centering
	\includegraphics[width=8cm]{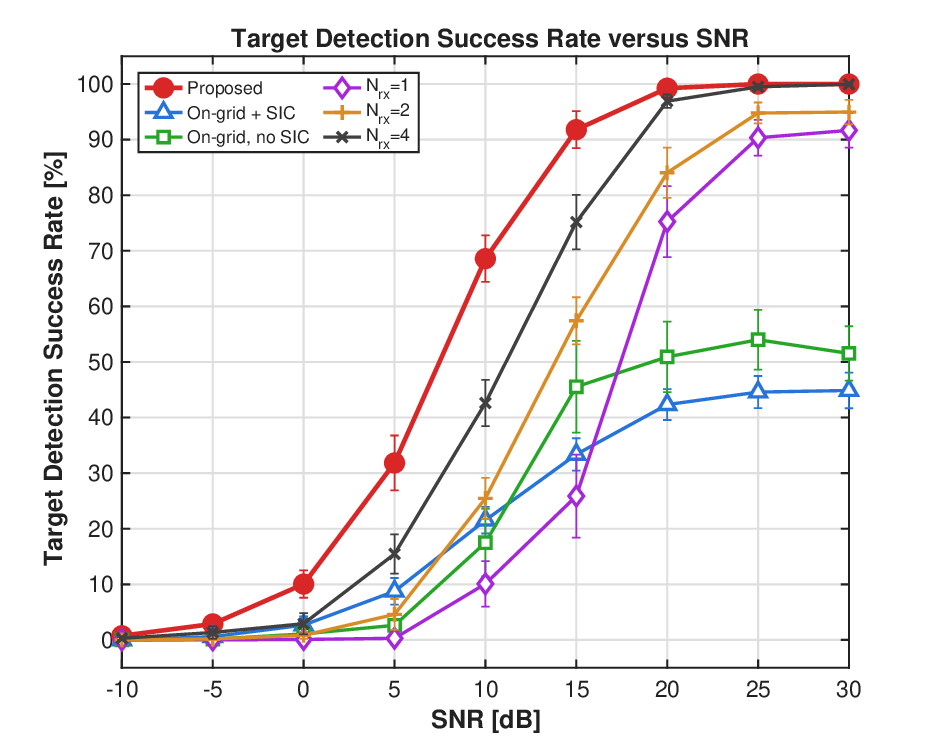}
	\caption{Target detection success rate of the proposed localization method and its off-grid, SIC, and receive-array ablations.}
	\label{Fig_SR_Ablation}
	\vspace*{-5mm}
\end{figure}

We evaluate the proposed localization method and five ablation variants, with all other simulation parameters unchanged.
The ``on-grid + SIC'' variant disables the local fine search and off-grid interpolation, while the ``on-grid, no SIC'' variant further disables residual cancellation and its associated candidate-acceptance test. 
Three receive-array variants retain the complete processing chain while reducing $N_{\rm rx}^{(\rm A)}$ to $1$, $2$, or $4$. 
A target is successfully detected when its 3D localization error is below $6$ m, and errors above $15$ m are capped at $15$ m when computing the RMSE. 
The vertical error bars represent two-sided $90\%$ Student-$t$ confidence intervals calculated over $15$ independent trajectory--noise sequences.
Figs.~\ref{Fig_RMSE_Ablation} and \ref{Fig_SR_Ablation} show that the complete localization method reaches the reliable-tracking regime at a lower SNR.
Its advantage over the on-grid variants demonstrates that local fine search and off-grid interpolation jointly alleviate grid mismatch and preserve the accuracy of residual reconstruction.
The ``on-grid + SIC'' variant does not outperform its no-SIC counterpart, indicating that coarse-grid mismatch can produce inaccurate echo reconstruction and thereby degrade subsequent processing. As $N_{\rm rx}^{(\rm A)}$ is reduced from $8$ to $4$, $2$, and $1$, the RMSE increases and the detection success rate consistently decreases, confirming the spatial-discrimination gain provided by the BS-A receive array.


\subsubsection{Localization Baseline Comparison}

{To evaluate the proposed method, we consider ROI-based cooperative maximum likelihood with successive interference cancellation (ROI-CML-SIC) and full-grid joint cooperative maximum likelihood (FG-JCML), both following the cooperative concentrated-likelihood objective in \cite{Pucci2025CooperativeML}, together with sparse recovery and refinement (SRR) adapted from \cite{Zhang2025DirectMTL}. ROI-CML-SIC adds 3D multi-target search, sequential LS gain refitting, and residual cancellation to the single-target cooperative ML position objective. FG-JCML evaluates the objective over the complete 3D coarse grid, applies NMS, jointly scores candidate pairs, and performs local coordinate refinement. SRR retains alternating direction method of multipliers (ADMM) recovery, continuous refinement, and successive cancellation, while the 3D multi-static adaptation adds group-sparse recovery, active-set screening, and joint gain refitting. All methods use identical observations, trajectories, DPI preprocessing, global coarse-grid support, target count, evaluation rules, and a common output tracker.}

\begin{figure}[!t]
	\captionsetup{font={footnotesize}, name = {Fig.}, labelsep = period}
	\centering
	\includegraphics[width=8cm]{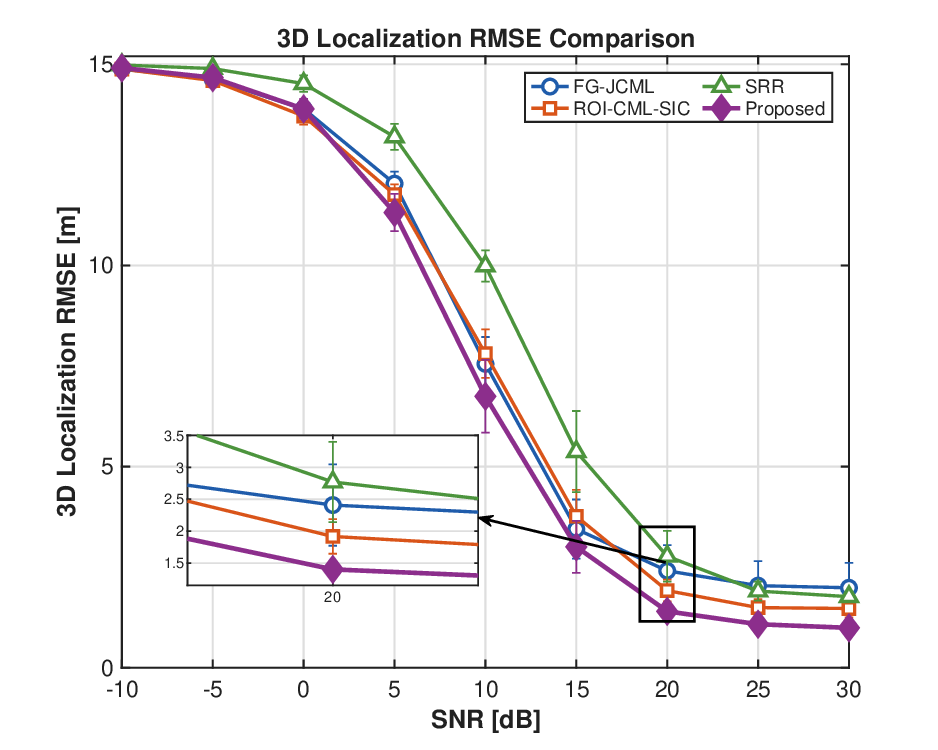}
	\caption{3D localization RMSE performance of the proposed method and representative localization baselines.}
	\label{Fig_RMSElocal}
	\vspace*{-5mm}
\end{figure}
\begin{figure}[!t]
	\captionsetup{font={footnotesize}, name = {Fig.}, labelsep = period}
	\centering
	\includegraphics[width=8cm]{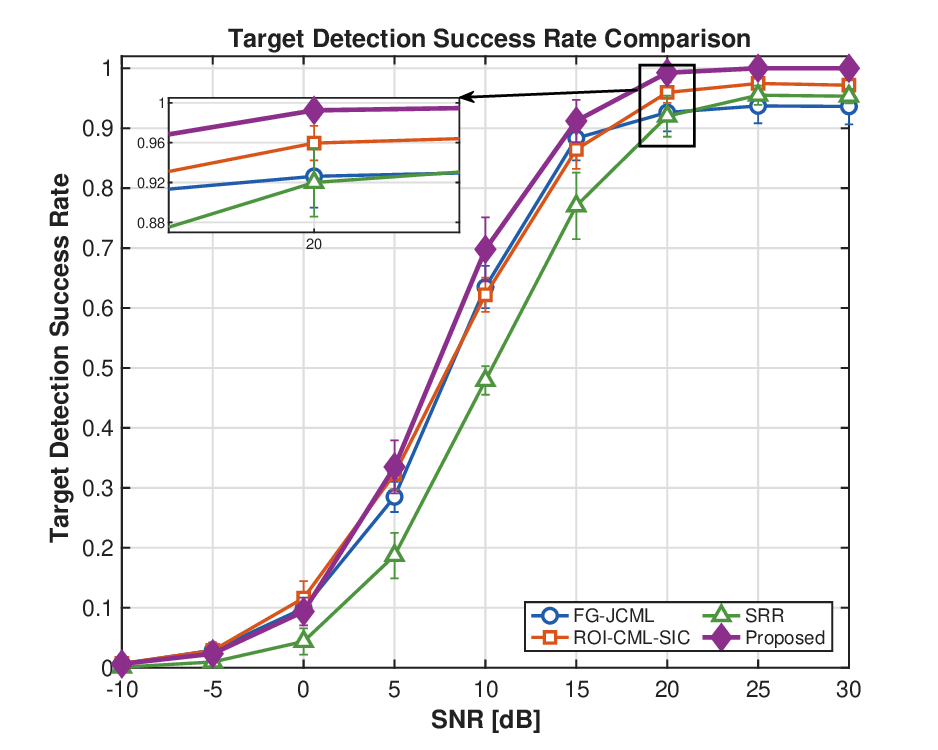}
	\caption{Target detection success rate of the proposed method and representative localization baselines.}
	\label{Fig_SRlocal}
	\vspace*{-5mm}
\end{figure}

{For complexity comparison, ROI-CML-SIC has pure-search cost $\mathcal C_{\rm ROI}^{\rm search}=\mathcal O\!\bigl(QN_{\rm fine}(N_{\rm c}+N_{\rm ROI})\bigr)$, while its joint gain refitting and residual reconstruction are additional estimation costs. For FG-JCML with $Q=2$, $L_{\rm c}$ and $H_{\rm c}$ count retained candidates and refined candidate pairs, $N_{\rm pre}$ and $N_{\rm ref}$ count grids per pre-refinement and coordinate update, and $J_{\rm coord}$ counts those updates, yielding $\mathcal C_{\rm FG}^{\rm search}=\mathcal O\!\bigl(N_{\rm fine}[N_{\rm c}+\allowbreak L_{\rm c}N_{\rm pre}+\allowbreak L_{\rm c}^{2}+\allowbreak H_{\rm c}J_{\rm coord}N_{\rm ref}]\bigr)$. For SRR, $A$ is the active-set size and $J_{\rm ref}$ is the refinement iteration count, giving $\mathcal C_{\rm SRR}^{\rm search}=\mathcal O\!\bigl(QN_{\rm fine}(N_{\rm c}+\allowbreak A+\allowbreak J_{\rm ref})\bigr)$. The term $\mathcal O\!\bigl(Q\sum_{i'\in\mathcal I}(A^3+\allowbreak J_{\rm ADMM}A^2)\bigr)$ is reported separately as ADMM recovery cost. Thus, ROI-CML-SIC repeats full-grid searches, FG-JCML adds candidate-pair processing, and SRR adds active-set screening. Stable tracking with the proposed method requires only $\mathcal O(QN_{\rm fine}N_{\rm loc})$. Fallback uses the smaller dimension $N_{\rm coa}$.}

{As shown in Figs.~\ref{Fig_RMSElocal} and \ref{Fig_SRlocal}, noise dominates the target echoes in the low-SNR region, resulting in unreliable localization and no consistent performance ordering. As the SNR increases, the proposed method achieves a lower average RMSE and a higher average target detection success rate.
This performance is consistent with the coordinated use of kinematic prediction, NMS-based multi-candidate screening, adaptive ROI expansion, off-grid refinement, residual verification, and SIC.
Multi-candidate screening and ROI expansion reduce coarse-search ambiguities, while off-grid refinement reduces spatial discretization errors and residual verification limits SIC mismatch. Together with the complexity analysis above, simulations demonstrate a favorable performance--complexity balance under the common observation model.}

\subsubsection{Robustness against Inter-BS Clock Offsets}
\begin{figure}[!t]
	\captionsetup{font={footnotesize}, name = {Fig.}, labelsep = period}
	\centering
	\includegraphics[width=8cm]{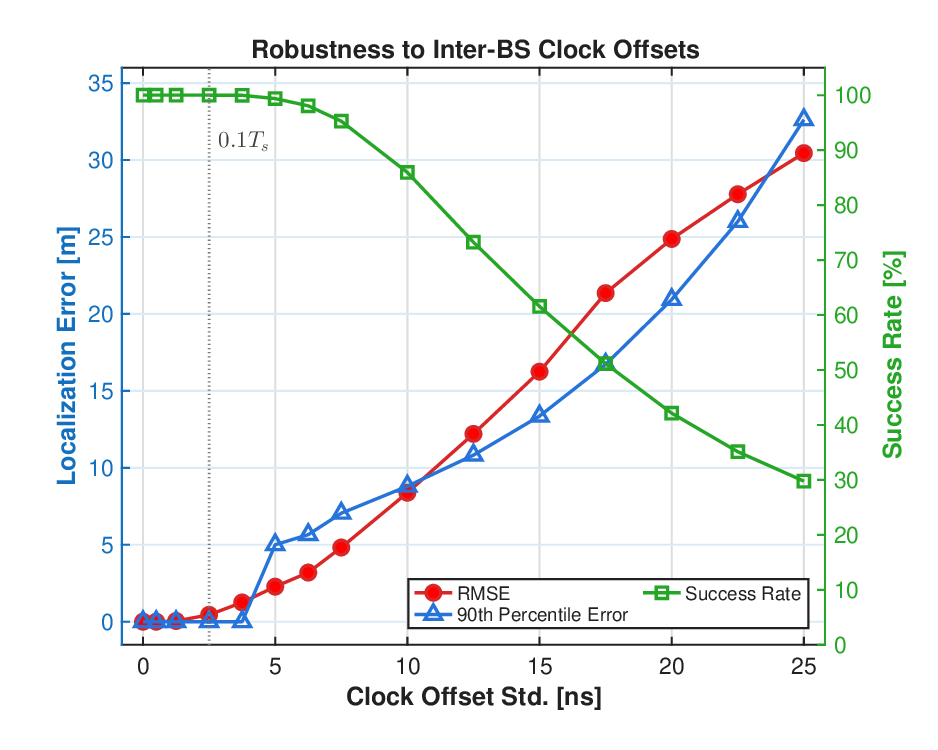}
	\caption{Multi-static cooperative localization performance versus the standard deviation of residual inter-BS clock offsets.}
	\label{Fig_clock_offset}
	\vspace*{-5mm}
\end{figure}

{
Although the proposed localization method is not a conventional TDOA algorithm, residual inter-BS clock offsets perturb the geometry-dependent delays used by its sensing dictionary. For an offset $\Delta t_i^{\rm{clk}}$ at the $i$-th receiving BS, the effective sample-domain delay is $\widetilde{\tau}_i(\mathbf{p})=\tau_i(\mathbf{p})+\Delta t_i^{\rm{clk}}/T_{\rm{s}}$. We set BS-A as the time reference with $\Delta t_{\mathrm{A}}^{\rm{clk}}=0$ and independently generate $\Delta t_i^{\rm{clk}}\sim\mathcal{N}(0,\sigma_{\rm{clk}}^2)$ for the other BSs. The observations contain these offsets, whereas the dictionary assumes ideal synchronization. At an SNR of $25~\mathrm{dB}$, Fig.~\ref{Fig_clock_offset} shows negligible degradation for $\sigma_{\rm{clk}}\leq0.1T_{\rm{s}}$. The 90th-percentile curve reports the 90th percentile of the 3D localization error and shows that tail errors grow with $\sigma_{\rm{clk}}$ alongside the RMSE. Larger offsets weaken the joint dictionary match, increasing the localization error and reducing the success rate, with pronounced degradation near half a sampling interval. Because this transition depends on the sampling rate, SNR, and BS-target geometry, dominant offsets should be calibrated, while fully asynchronous operation requires joint clock-position estimation \cite{wang2020tdoa}.
}

\vspace{-5mm}
\section{Conclusion}
This paper proposed a UW-OCDM ISAC framework for low-altitude UAV networks, jointly supporting high-mobility communication and multi-target sensing.
	At the waveform level, UW embedding established the circular convolution structure required for low-complexity FDE, while the quadratic phase rotation of OCDM enhanced Doppler resilience and mitigated the repetitive correlation ambiguity of UW-OFDM.
	Additionally, the known UW supported time synchronization, Doppler estimation and compensation, and sparse ST CE. 
	For cooperative sensing, the UW was also utilized for multi-static UAV localization. 
	Since the UW and system parameters were pre-stored at the cooperating BSs, sensing dictionaries were constructed locally without exchanging full payload data, thereby reducing transmit-reference sharing overhead.
	Combined with DPI suppression and hierarchical off-grid processing, the proposed localization method achieved robust multi-UAV localization under grid ambiguity and near-far interference.
	Numerical results demonstrated improved communication reliability and multi-target localization accuracy, confirming that the proposed UW-driven design effectively supported both robust communication reception and cooperative sensing through their respective processing chains.

Building on the demonstrated benefits of the UW-driven processing chain, future work will investigate its extension to other candidate waveforms and more general ISAC scenarios. Such extensions will require waveform-specific redesign of the UW embedding mechanism, parameter configuration, and receiver processing. The framework will also be further developed for distributed, multi-user, and learning-assisted ISAC systems under more practical synchronization, hardware, and environmental conditions.

\bibliography{refs}

\end{document}